\documentclass[floatfix,showpacs,preprintnumbers,amsmath,amssymb,aps,twocolumn,superscriptaddress]{revtex4}

\usepackage{amsfonts}
\usepackage[pdftex]{graphicx}
\usepackage{type1cm}
\usepackage{color}
\usepackage{bm}
\newcommand{\mat}[1]{\mathsf{#1}}
\newcommand{\figref}[1]{Fig.~\ref{#1}}
\newcommand{\e}[1]{\text{e}^{#1}}
\newcommand{\cmplxi}{\text{i}}
\newcommand{\diffd}{\text{d}}
\newcommand{\tr}{\operatorname{Tr}}
\renewcommand{\vec}[1]{\mathbf{#1}}
\newcommand{\punc}[1]{\,#1}
\newcommand{\neweqnline}{\nonumber\\}
\newcommand{\secref}[1]{Sec.~\ref{#1}}
\newcommand{\eqnref}[1]{Eqn.~(\ref{#1})}
\newcommand{\vecgrk}[1]{\boldsymbol{#1}}

\newcommand\ksi{\xi}

\begin{document}

\title{First principles calculation of conductance and current flow through low-dimensional superconductors}
\author{G.J.~Conduit}
\affiliation{Department of Physics, Ben Gurion University, Beer Sheva 84105, Israel}
\author{Y.~Meir}
\affiliation{Department of Physics, Ben Gurion University, Beer Sheva 84105, Israel}

\date{\today}

\begin{abstract}
We present a novel formulation to calculate transport through
disordered superconductors connected between two metallic leads. An
exact expression for the current is derived, and is applied to a
superconducting sample described by the negative-$U$ Hubbard model. A
Monte Carlo algorithm that includes thermal phase and amplitude
fluctuations of the superconducting order parameter is employed, and a
new efficient algorithm is described. This improved routine allows
access to relatively large systems, which we demonstrate
by applying it to several cases, including superconductor-normal
interfaces and Josephson junctions.  The effects of decoherence and
dephasing are shown to be included in the formulation, which allows the
unambiguous characterization of the Kosterlitz-Thouless transition in
two-dimensional systems and the calculation of the finite resistance due to vortex
excitations in quasi one-dimensional systems. Effects of magnetic
fields can be easily included in the formalism, and are demonstrated
for the Little-Parks effect in superconducting cylinders. Moreover,
the formalism enables us to map the local super and normal currents,
and the accompanying electrical potentials, which we use to pinpoint
and visualize the emergence of resistance across the
superconductor-insulator transition.
\end{abstract}


\maketitle

\section{Introduction}

Chief amongst the remarkable effects observed in superconductors is
their eponymous perfect conductivity. Within BCS theory \cite{BCS},
where superconductivity arises due to pairing between electrons, the
effects of temperature $T$, magnetic field $B$, and disorder are well
understood: as the pairing amplitude is suppressed by these physical
parameters, the system becomes normal, and attains a finite
resistance. For low-dimensional systems, on the other hand, it has
been long understood that phase fluctuations of the pairing amplitude
play a major role in the loss of perfect
conductance~\cite{Halperin}. In two-dimensional systems, for example,
it has been demonstrated~\cite{BKT} that as the temperature increases
there is a critical temperature $T_{\text{KT}}$ where vortices and
anti-vortices unbind and proliferate through the system, leading to
the loss of global phase coherence and superconductivity, even though
the pairing amplitude remain finite. Indications of such a
Berezinsky-Kosterlitz-Thouless (BKT) transition have been observed in
Josephson-junction arrays~\cite{KTarray}, in superconducting (SC) thin
films~\cite{Yazdani93}, and possibly in high-$T_{\text{c}}$
cuprates~\cite{KTinHighTc}.

In recent years there has been a reinvigoration of research into
low-dimensional superconductors. This has been motivated by intriguing
experimental observations of electronic transport through disordered
SC thin films, such as a huge magnetoresistance peak~\cite{MR} and a
``super-insulator'' phase~\cite{SI}, and by the technological progress
in producing two-dimensional superconductors in the interface between
two oxides~\cite{LaSr} and in making ultra-thin cuprate
superconductors~\cite{UltraThin}.  Many of these observations are not
yet satisfactory explained, chiefly because there is no theory that
can calculate the current, even numerically, through a disordered
superconductor, based on a microscopic model.

The calculation of the resistance within the BCS picture, usually
based on the Bogoliubov-de Gennes (BdG) mean-field approach, is
straightforward.  Blonder, Tinkham, and Klapwijk
(BTK)~\cite{Blonder82} studied the reflectance and transmission at a
metal-superconductor junction, and an analogous study was performed at
superconductor-metal-superconductor junctions~\cite{Kulik70}.  Similar
approaches~\cite{resistance} utilized the Buttiker-Landauer
picture~\cite{Landauer57,Buttiker86} for non-interacting Cooper pairs
to study scattering through a SC region. (A difficulty with the direct
application of the BdG formalism is the non-conservation of charge,
which can be overcome by studying a normal ring containing a SC
segment~\cite{EntinWohlman08}.) An alternative approach near to the
BCS critical temperature is to use a scaling assumption for the
conductivity~\cite{Dorsey91}. The current through diffusive normal
metal-superconductor structures has also been calculated using a
Keldysh scattering matrix theory~\cite{Stoof95}. All these approaches
neglect phase fluctuations so cannot be used to study two-dimensional
superconductors that exhibit a BKT-like transition at low
temperatures.

The resistance of low-dimensional superconductors can also be
calculated using phenomenological models. The conductivity of uniform
systems can be probed analytically by studying phase slips across the
sample within the Ginzburg-Landau
approach~\cite{Langer67,McCumber70}. Thermally excited phase slips
explain both non-linear conductivity and vortex creep induced
resistance~\cite{Halperin79}, whilst quantum activated phase slips can
drive SC wires insulating~\cite{Buchler04}.  However, phenomenological
calculations are neither underpinned by a microscopic model nor
include Coulomb repulsion or disorder except for the introduction of a
phenomenological normal state resistance.

Here we develop a new formalism to calculate the current through a
superconductor taking into account phase fluctuations in the presence
of disorder, finite $T$ and $B$, and Coulomb repulsion.  The approach
we detail here is based on the Landauer-Buttiker scheme
\cite{Landauer57,Buttiker86}, where one attaches metallic leads to the
sample, and then calculate its conductance.  The lead-superconductor
tunneling barriers ensure that the conductance of the system is always
finite, even in the SC phase.  A previous attempt using the quantum
Monte Carlo approach to calculate current in disordered systems
employed the fluctuation-dissipation theorem via the current-current
correlation function~\cite{Trivedi96}.

\begin{figure}
 \includegraphics[width=0.95\columnwidth]{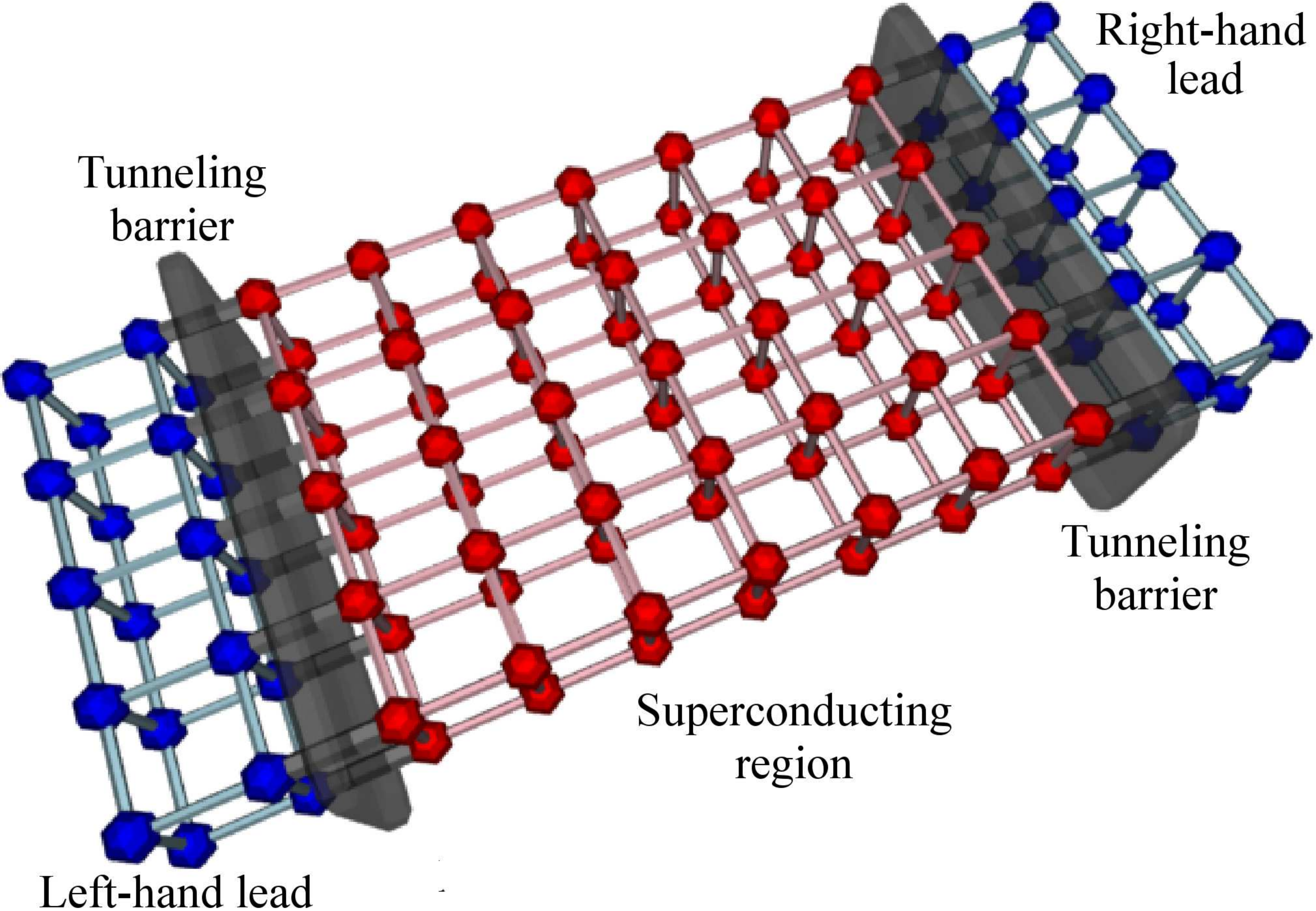}
 \caption{(Color online) A schematic of the experimental setup within
   the negative-$U$ Hubbard model. The left and right-hand metallic leads are shown in
   blue, from which electrons can tunnel through the barriers shown by the gray links into
   the central SC region which is shown in red.}
 \label{fig:Setup}
\end{figure}

The Landauer formula~\cite{Landauer57,Buttiker86} is a widely adopted
method to calculate the current through a mesoscopic sample that
contains non-interacting particles. Meir and Wingreen \cite{Meir92}
have generalized the formula to produce an exact expression for the
current through any interacting region attached to non-interacting
leads, which has been successfully applied to a wide range of
systems. Following this approach, we partition the system into the
three parts shown in \figref{fig:Setup}: the left-hand lead, the
central interacting region, here a superconductor, and the right-hand
lead. In the leads the natural particle basis set are electrons, and
in the sample the natural basis set are Bogoliubons. To circumvent
this mismatch of particle basis sets we reformulate the Meir-Wingreen
formula in a Bogoliubon basis set to derive an exact expression for
the current flow through a possibly SC region, attached to two
metallic leads.

Having derived a general, exact formula, the SC region is then modeled
by a generalized negative-$U$ Hubbard model based on the
lattice shown in \figref{fig:Setup}. Introducing two local
auxiliary fields (which reduce to the local density and gap at zero
temperature), we decouple the interacting fermions. While the
conductance formula is exact, in order to evaluate correlation
functions we neglect quantum fluctuations, and integrate numerically
over the thermal fluctuations of the auxiliary parameters
\cite{Mayr,Erez2010} using a Monte Carlo method.  A significant
advantage of the formalism is that it allows us to construct current
and potential maps of the system. These allow us to diagnose the
microscopic features that increase the resistance of the sample. This
paper details the new procedure and presents a number of applications
of the formalism for simple systems, where one can compare with
existing theories.

The paper is organized as follows: in \secref{sec:AnalyticalFormalism}
we first derive an exact expression for the current through a SC
region. Using this expression, in
\secref{sec:NumericalProcedure} we describe how the current can be
calculated numerically and outline improvements to the auxiliary field
approach that allows us to study large systems. Having developed the
new formula for the current and accompanying computational tool, it is
vital to carefully test it against a series of known
results. Therefore, in \secref{sec:LengthDependence} we study
superconductor-normal interfaces in clean systems, and compare with
the BTK transmission formulae, while in
\secref{sec:JosephsonJunction} we study the temperature dependent current in a Josephson
junction. In \secref{sec:SelfEnergy} we describe how effects of
decoherence and dephasing are manifested in the formalism. We
investigate the temperature dependence of resistance in
\secref{sec:KosterlitzThoulessTransition} in which we uncover the
temperature dependence of the resistance and the nonlinear $I-V$
behavior that characterizes the BKT transition in two dimensions and
vortex excitations in quasi one-dimensional systems.  We then, in
\secref{sec:LittleParksEffect}, apply an external magnetic field to
probe the Little-Parks effect. Finally, we demonstrate how to
construct current and potential maps for the system, and use them to
study the microscopic behavior at the superconductor-insulator
transition in \secref{sec:CurrentMaps}. The details of the analytical
derivation and the numerical procedure are described in the appendices.

\section{Analytical Derivation}\label{sec:AnalyticalFormalism}
\subsection{Current Formula}

To calculate the current for interacting particles we start with the
general formula for the current~\cite{Meir92} through an interacting
region, connected between two non-interacting leads
\begin{align}
 J=\frac{\cmplxi e}{2h}\sum_\sigma
 \int\diffd\epsilon\Bigl[&\tr
\left\{\left(f_{\text{L}}(\epsilon)\mat{\Gamma}^{\text{L}}-f_{\text{R}}(\epsilon)\mat{\Gamma}^{\text{R}}\right)
\left({\cal G}_{\sigma}^{\text{r}}-{\cal G}_{\sigma}^{\text{a}}\right)\right\}\neweqnline
  +&\tr\left\{(\mat{\Gamma}^{\text{L}}-\mat{\Gamma}^{\text{R}}){\cal G}_{\sigma}^{<}\right\}\Bigr]\punc{.}
\label{eqn:MeirWingreen}
\end{align}
Here
$f_{\chi}(\epsilon)\equiv[\exp(\beta(\epsilon-\mu_{\chi}))+1]^{-1}$
with $\chi\in\{\text{L},\text{R}\}$ is the Fermi distribution of the
left (L) and right-hand (R) leads that are held at chemical
potentials $\mu_{\chi}$ and reduced temperature
$\beta\equiv1/k_{\text{B}}T$ (where $k_B$ is the Boltzmann constant). The
imposed potential difference $eV\equiv\Delta\mu=\mu_{\text{L}}-\mu_{\text{R}}$
between the leads drives the current $J$ through the system.  The
integral is over all electronic energies $\epsilon$.
$\mat{\Gamma}^{\chi}_{ij}\equiv2\pi\sum_{a\in\chi}\rho_{a}(\epsilon)Y_{ai}Y_{aj}^{*}$
for channels $a$ in lead $\chi$, and $Y_{a,i}$ is the tunneling matrix
element from channel $a$ in the the lead to site $i$ in the
sample. Finally, ${\cal G}_{ij\sigma}^{\text{r}}$, ${\cal
G}_{ij\sigma}^{\text{a}}$, and ${\cal G}_{ij\sigma}^{<}$ are the electronic
retarded, advanced, and lesser Green functions (in the site basis) for electrons of spin
$\sigma$ in the sample calculated in the presence of the leads.

\eqnref{eqn:MeirWingreen} is exact, and captures, via the
electronic Green function ${\cal G}$, all the processes that can
transfer an electron through the system. When the intermediate regime
has SC correlations, some of these processes involve Andreev
scattering -- absorption of an electron pair by the condensate and a
propagation of the remaining hole. To expose these processes, it is
convenient to transform from the electron basis set
$(c_{i\sigma}^{\dagger},c_{i\sigma})$ with site index $i$ into the
Bogoliubov basis set $(\gamma_{n\sigma}^{\dagger},\gamma_{n\sigma})$,
using the Bogoliubov-de Gennes relations
$c_{i\sigma}=\sum_{n}u_{i}(n)\gamma_{n\sigma}-\sigma
v_{i}^{*}(n)\gamma_{n-\sigma}^{\dagger}$ (at present $u_{i}$ and
$v_{i}$ are arbitrary, except for the unitarity condition, but later on they will be determined
by the actual Hamiltonian that will be used for the SC region). The
Green functions transform from the electron basis ${\cal G}_{\sigma}$
into the energy basis set of Green functions
$\{\mat{G}_{\sigma}^{>},\mat{G}_{\sigma}^{<}\}$ and the family of
anomalous Green functions
$\mat{H}^{>}_{\sigma}(m,n)=-\cmplxi\langle\gamma_{m-\sigma}^{\dagger}\gamma_{n\sigma}^{\dagger}\rangle$,
$\mat{H}^{<}_{\sigma}(m,n)=\cmplxi\langle\gamma_{n-\sigma}^{\dagger}\gamma_{m\sigma}^{\dagger}\rangle$,
$\bar{\mat{H}}^{>}_{\sigma}(m,n)=-\cmplxi\langle\gamma_{m-\sigma}\gamma_{n\sigma}\rangle$,
and
$\bar{\mat{H}}^{<}_{\sigma}(m,n)=\cmplxi\langle\gamma_{n-\sigma}\gamma_{m\sigma}\rangle$
according to
\begin{align}
 &{\cal G}_{\sigma}^{\text{r}}(i,j)-{\cal G}_{\sigma}^{\text{a}}(i,j)=
{\cal G}_{\sigma}^{>}(i,j)-{\cal G}_{\sigma}^{<}(i,j)\neweqnline
 &=\vec{u}_{i}\left(\mat{G}_{\sigma}^{>}-\mat{G}_{\sigma}^{<}\right)\vec{u}_{j}^{*}+\vec{v}_{i}\left(\mat{G}_{-\sigma}^{>}-\mat{G}_{-\sigma}^{<}\right)\vec{v}_{j}^{*}\neweqnline
 &-\sigma\vec{v}_{i}^{*}\left(\mat{H}_{\sigma}^{>}-\mat{H}_{\sigma}^{<}\right)\vec{u}_{j}^{*}-\sigma\vec{u}_{i}\left(\bar{\mat{H}}_{-\sigma}^{>}-\bar{\mat{H}}_{-\sigma}^{<}\right)\vec{v_{j}}\punc{,}
\end{align}
and
\begin{align}
 {\cal G}_{\sigma}^{<}(i,j)\!=\!
 \vec{u}_{j}^{*}\mat{G}^{<}_{\sigma}\vec{u}_{i}^{*}\!-\!
 \vec{v}_{j}\mat{G}^{>}_{-\sigma}\vec{v}_{i}^{*}\!+\!
 \sigma\vec{u}^{*}_{j}\mat{H}_{\sigma}^{>}\vec{v}^{*}_{i}\!-\!
 \sigma\vec{v}_{j}\bar{\mat{H}}_{-\sigma}^{<}\vec{u}_{i}\!\punc{.}
\end{align}
Solving for the
Green functions across the system in the presence of the leads
(Appendix A), leads to the final, exact result
\begin{align}
 &J=\frac{e}{h}\sum_{\sigma}\int\diffd\epsilon[f_{\text{L}}(\epsilon)-f_{\text{R}}(\epsilon)]\times\neweqnline
\tr\Bigl[&(\mat{\Gamma}_{\mat{u}^{*}\mat{u}}^{\chi}+\mat{\Gamma}_{\mat{v}^{*}\mat{v}}^{\chi})\mat{G}_{\sigma}^{\text{a}}(\mat{\Gamma}_{\mat{u}\mat{u}^{*}}^{-\chi}-\mat{\Gamma}_{\mat{v}\mat{v}^{*}}^{-\chi})\mat{G}_{\sigma}^{\text{r}}\neweqnline
           +&(\mat{\Gamma}_{\mat{u}\mat{v}}^{\chi}-\mat{\Gamma}_{\mat{v}\mat{u}}^{\chi})\mat{G}_{\sigma}^{\text{a}}\mat{\Gamma}_{\mat{v}^{*}\mat{u}^{*}}^{-\chi}\mat{H}_{\sigma}^{\text{r}}+(\mat{\Gamma}_{\mat{u}^{*}\mat{v}^{*}}^{\chi}-\mat{\Gamma}_{\mat{v}^{*}\mat{u}^{*}}^{\chi})\mat{G}_{\sigma}^{\dagger\text{a}}\mat{\Gamma}_{\mat{u}\mat{v}}^{-\chi}\mat{H}_{\sigma}^{\dagger\text{r}}\neweqnline
           +&\sigma\mat{\Gamma}_{\mat{u}\mat{u}^{*}}^{\chi}\mat{H}_{\sigma}^{\text{a}}(\mat{\Gamma}_{\mat{v}^{*}\mat{u}^{*}}^{-\chi}\!\!-\!\mat{\Gamma}_{\mat{u}^{*}\mat{v}^{*}}^{-\chi})\mat{G}_{\sigma}^{\text{r}}\!+\!\sigma\mat{\Gamma}_{\mat{v}\mat{v}^{*}}^{\chi}\mat{H}_{\sigma}^{\dagger\text{a}}(\mat{\Gamma}_{\mat{v}\mat{u}}^{-\chi}\!\!-\!\mat{\Gamma}_{\mat{u}\mat{v}}^{-\chi})\mat{G}_{\sigma}^{\dagger\text{r}}\neweqnline
           +&\sigma(\mat{\Gamma}_{\mat{u}\mat{u}^{*}}^{\chi}+\mat{\Gamma}_{\mat{v}\mat{v}^{*}}^{\chi})(\mat{H}_{\sigma}^{\text{a}}\mat{\Gamma}_{\mat{v}^{*}\mat{u}^{*}}^{-\chi}\mat{H}_{\sigma}^{\text{r}}+\mat{H}_{\sigma}^{\dagger\text{a}}\mat{\Gamma}_{\mat{u}\mat{v}}^{-\chi}\mat{H}_{\sigma}^{\dagger\text{r}})\Bigr]\punc{,}
\label{eqn:GeneralEquationForCurrent}
\end{align}
where
$\mat{\Gamma}^{\chi}_{\mat{u}\mat{v}}(m,n)=2\pi\sum_{i,j,a\in\chi}\rho_{a}(\epsilon)
Y_{ai}Y_{aj}\mat{u}_{i}(m)\mat{v}_{j}(n)$ is now in the transformed
basis set.  This is written in a form describing transmission from the
left-hand side to the right-hand side of the sample. We will show in
\secref{sec:LengthDependence} that it therefore exposes
the rise of resistance due to the suppression of correlations between
the left and right-hand sides of the superconductor.

We note that deep in the SC regime where the SC gap obeys $\Delta\gg Y$,
and in the case where the leads inject electrons within the gap such that
$eV<2\Delta$, we can make a perturbative expansion in
small tunneling $Y$. This yields the simple expression for the current
\begin{align}
 J=&\frac{e}{h}\sum_{\sigma}\int\diffd\epsilon[f_{\text{L}}(\epsilon)-f_{\text{R}}(\epsilon)]\times\neweqnline
 &\tr\left[(\mat{\Gamma}_{\mat{u}^{*}\mat{u}}^{\chi}+\mat{\Gamma}_{\mat{v}^{*}\mat{v}}^{\chi})\tilde{\mat{G}}_{\sigma}^{\text{a}}(\mat{\Gamma}_{\mat{u}\mat{u}^{*}}^{-\chi}-\mat{\Gamma}_{\mat{v}\mat{v}^{*}}^{-\chi})\tilde{\mat{G}}_{\sigma}^{\text{r}}\right]\punc{.}
 \label{eqn:PerturbativeEquationForCurrent}
\end{align}
Here $\tilde{\mat{G}}$ are the Green functions calculated in the
absence of the leads.  This equation has direct $Y^{4}$ dependence on
the tunneling matrix element, with neglected higher order
contributions of order $\sim(Y/\Delta)^{6}$, as it describes Cooper
pairs tunneling through the contact barrier.  We shall show later that
this contribution is precisely what is predicted for the
current~\cite{Lambert91} according to the BTK
formula~\cite{Blonder82}, and notably, as the leads inject electrons
only into the gap, there is no normal current, but only Andreev
processes allow the flow of current. The perturbative form
\eqnref{eqn:PerturbativeEquationForCurrent} offers two
important computational advantages. Firstly, it is considerably less
resource intensive to calculate as the Green functions are diagonal so
it does not demand summations over separate variables. Secondly, it
does not require the expensive matrix inversion embodied in
\eqnref{eqn:MatrixEquationForRetardedAndAdvancedGreensFunctions} to
find the  general equation for the current. Due to its usefulness
we also note that an analogous expression can be derived for the normal current
when injecting electrons outside of the gap
\begin{align}
 J=\frac{e}{h}\sum_{\sigma}\int\diffd\epsilon\tr[&f_{\text{L}}(\epsilon)(\mat{\Gamma}_{\mat{u}^{*}\mat{u}}^{\chi}+\mat{\Gamma}_{\mat{v}^{*}\mat{v}}^{\chi})\neweqnline
  -&f_{\text{R}}(\epsilon)(\mat{\Gamma}_{\mat{u}^{*}\mat{u}}^{\chi}+\mat{\Gamma}_{\mat{v}^{*}\mat{v}}^{\chi})]\Im\mat{G}_{\sigma}^{\text{r}}\punc{,}
 \label{eqn:PerturbativeEquationForNormalCurrent}
\end{align}
where $\Im$ stands for the imaginary part. Since this term represents
the normal current, it has a direct $Y^{2}$ dependence on the
tunneling matrix element. Though they offer a considerable
computational advantage, these perturbative formulae cannot be used on
the border of the superconductor-insulator transition where the
superconductor gap breaks down and $\Delta<Y$. Therefore, unless
specified, we use the full expression for the current,
\eqnref{eqn:GeneralEquationForCurrent}, in our numerical calculations.

\subsection{Current and voltage maps}

\eqnref{eqn:PerturbativeEquationForNormalCurrent}, with a coefficient
of $Y^{2}$, describes the normal current that enters and leaves the
system as single electrons, whereas
\eqnref{eqn:PerturbativeEquationForCurrent} with a coefficient of
$Y^{4}$ corresponds to a tunneling supercurrent. However, the normal
and supercurrent can interchange inside the sample. In order to
understand the microscopics behind phenomena in the disordered
superconductor it is vital that we can probe the spatial distribution
of the current as it switches in nature through the sample. Therefore, here we
extend our formalism to map out the flow of current within the sample.
To calculate the current distribution map we use the general
expression for the current crossing a single
bond~\cite{Caroli71,Cresti03} from site $i$ to $j$
\begin{align}
 J_{ij}=\frac{2e}{h}\sum_{\sigma}\int\frac{\diffd\epsilon}{2\pi}\left[t_{ij}{\cal G}_{\sigma}^{<}(j,i)-t_{ji}{\cal G}_{\sigma}^{<}(i,j)\right]\punc{.}
\end{align}
Transforming again into the diagonalized basis, the local current is
\begin{align}
 J_{ij}=&\frac{2e}{h}\sum_{\sigma}\!\int\!\frac{\diffd\epsilon}{2\pi}\tr\biggl
\{\!\left[\mat{\Lambda}^{ij}_{\mat{u}^{*}\mat{u}}\!-\!\mat{\Lambda}^{ij}_{\mat{v}^{*}\mat{v}}\right]\!\mat{G}_{\sigma}^{<}\!-\!\left[\mat{\Lambda}^{ij\text{T}}_{\mat{u}\mat{u}^{*}}\!-\!\mat{\Lambda}^{ij\text{T}}_{\mat{v}\mat{v}^{*}}\right]\!\mat{G}_{\sigma}^{<}\neweqnline
 +&\sigma\left[\mat{\Lambda}^{ij\text{T}}_{\mat{v}^{*}\mat{u}^{*}}-\mat{\Lambda}^{ij}_{\mat{u}^{*}\mat{v}^{*}}\right]\mat{H}_{\sigma}^{<}+\sigma\left[\mat{\Lambda}^{ij\text{T}}_{\mat{u}\mat{v}}-\mat{\Lambda}^{ij}_{\mat{v}\mat{u}}\right]\bar{\mat{H}}_{\sigma}^{<}\biggr\}\punc{,}
\end{align}
where
$\mat{\Lambda}^{ij}_{\vec{u}\vec{v}}(m,n)=t_{ij}\vec{u}_{i}(m)\vec{v}_{j}(n)$. As before, the normal $\mat{G}^{<}$
and anomalous Green functions
$\mat{H}^{<}$ are calculated in \eqnref{eqn:EqnForGLesser} in the
presence of the leads. Moreover we note that the current comes in two
flavors, the contribution to the current from the normal Green
function $\mat{G}^{<}$ is associated with the normal current and that
from the anomalous Green function $\mat{H}^{<}$ gives the Cooper pair
current. In \secref{sec:CurrentMaps} we verify that this intersite
current yields the correct net conservation of charge.

To provide an additional probe into the nature of the
superconductor-insulator transition we extend the formalism to map the
local chemical potentials across the sample. This should reveal any
weak links and the location of the sources of
resistance in a sample. To determine the local effective potential at
a specific site we add a weak link from that site to a third lead (a
``tip'').  The tunneling current from the tip into the sample is then
calculated, and the chemical potential of the tip adjusted until that
current flow is zero. This chemical potential thus corresponds to the
effective local chemical potential in that site. To calculate the
current flow into the tip we first evaluate the full Green functions
$\mat{G}$ in the sample in the presence of voltage drop between the
left and right reservoirs
\eqnref{eqn:MatrixEquationForRetardedAndAdvancedGreensFunctions} but
without the tip. We then use the perturbative formula for the current,
\eqnref{eqn:PerturbativeEquationForCurrent}, but with one lead
representing the left/right hand leads, and the other the perturbative
tip. This process is repeated for each site in the sample (due to the
perturbative nature of the tip, this calculation can be done
simultaneously for all sites). In
\secref{sec:CurrentMaps} we demonstrate how maps of the potential
can expose weak links in the sample and help diagnose the microscopic
mechanisms that give rise to resistance.

\section{Model and Numerical procedure}\label{sec:NumericalProcedure}

In the previous section we have developed an exact formula for the
current through an arbitrary intermediate region, which may include SC
correlations. We now use a specific model to describe this SC region
-- the negative-$U$ Hubbard model, a lattice model that includes
on-site attraction, and may include disorder, orbital and Zeeman
magnetic fields, and even long-range repulsive interaction (which we
will not deal with in this paper). The Hamiltonian is
\begin{align}
 \hat{H}_{\text{Hubbard}}&\!=\!\sum_{i,\sigma}\epsilon_{i\sigma}c^{\dagger}_{i\sigma}c_{i\sigma}\!
 -\sum_{i}U_{i}c^{\dagger}_{i\uparrow}c^{\dagger}_{i\downarrow}c_{i\downarrow}c_{i\uparrow}
 \neweqnline &-\!\!\!\sum_{\langle i,j\rangle,\sigma}\!\left(t_{ij}c^{\dagger}_{i\sigma}c_{j\sigma}+t_{ij}^{*}c^{\dagger}_{j\sigma}c_{i\sigma}\right)
 \punc{,}
 \label{eq:Hubbard}
\end{align}
where $\epsilon_{i\sigma}$ is the on-site energy, $t_{ij}$ the hopping
element between adjacent sites $i$ and $j$, and $U_{i}$ is the onsite
two-particle attraction, taken to be uniform, $i$-independent, in this paper. An
orbital magnetic field can be incorporated into the phases of the
hopping elements $t_{ij}$, while a Zeeman field splits the
spin-dependent on-site energies $\epsilon_{i\sigma}$. In this paper we
will only deal with orbital fields. To account for disorder,
$\epsilon_{i}$ will be drawn from a Gaussian distribution with
characteristic width $W$. The inter-site spacing is $a$.
Unlike, for example, the disordered
\emph{XY} model, the negative-$U$ Hubbard model can lead to a BCS
transition, a BKT transition, or to a percolation transition, and thus
this choice is general enough not to limit a priori the underlying
physical processes. Importantly, the model includes the fermionic
degrees of freedom which may be relevant to some of the experimental
observations.

Calculation of correlation functions, for example the Green functions
that enter the current formula, require thermal averages. To perform the thermal average
we need to decouple the quartic interaction term so
we employ the exact Hubbard-Stratonovich transformation
\begin{align}
 &\e{-\int_0^\beta\diffd\tau\sum_{i}U_{i}c_{i\uparrow}^\dag c_{i\downarrow}^\dag c_{i\downarrow}c_{i\uparrow}}=\neweqnline
 &\int\mathcal{D}\Delta\mathcal{D}\bar{\Delta}\e{-\int_0^\beta\diffd\tau \sum_i{\frac{-\left|\Delta_i(\tau)\right|^2}{U_{i}}+\Delta_i(\tau)c_{i\uparrow}^\dag c_{i\downarrow}^\dag+\bar{\Delta}_i(\tau)c_{i\downarrow}c_{i\uparrow}}},
\label{eq:HS}
\end{align}
which is basically a Gaussian integration ($\mathcal{D}\Delta\equiv
\Pi_{\tau,i}\diffd\Delta_i(\tau)$, where the product runs over all
times and all sites). Note that the field $\Delta_i(\tau)$ is just an
integration variable that decouples the two-body term in the
superconducting channel, and should not be confused with
$|U_{i}|\langle c_{i\downarrow}c_{i\uparrow}\rangle$. Similarly, one
introduces the integration fields $\rho_{i\sigma}(\tau)$, that couple
to the spin density~\cite{Altland06} $\langle c_{i\sigma}^\dag
c_{i\sigma}\rangle$, and leads to an additional term $-\sum_{i,\sigma}
|U_{i}|\rho_{i-\sigma}(\tau) c_{i\sigma}^\dag c_{i\sigma}$ in the
action (which, in the mean-field approximation gives rise to the
Hartree-Fock contribution) \footnote{If we consider the auxiliary
fields in momentum space, a complete summation over momentum
contributions from both $\Delta_{\vec{q}}$ and $\rho_{\vec{q}}$ would
lead to a double counting of the interaction term. However, in the
Monte Carlo calculation we sample just the low energy $q\ll
k_{\text{F}}$ fluctuations in each decoupling channel that will give
orthogonal contributions to the interaction term and avoid any double
counting~\cite{DePalo99}. In practice it was found that averaging over
fluctuations in the $\Delta$ field was important, and drove for
example a Kosterlitz Thouless transition. However, the average over
fluctuations in the $\rho$ field made only a negligible quantitative
change to the results.}.  Decoupling in both the $\Delta$ and $\rho$
channels not only provides access to both soft degrees of freedom, but
also the saddle point solution gives the standard mean-field results
for those fields, and furthermore guarantees that the action expanded
to Gaussian order corresponds to the random phase
approximation~\cite{DePalo99}.

The Hubbard-Stratonovich transformation (\ref{eq:HS}) is exact. Since
our main interest lies in thermal effects, for example the thermal BKT
phase transition, or thermal activation of vortices, we now neglect
quantum fluctuations (the $\tau$ dependence of $\Delta$). One can now
write the partition function for the Hubbard model as
\cite{Mayr,Erez2010}
\begin{equation}
 \mathcal{Z}=\tr\left[\e{-\beta\hat{H}_{\text{Hubbard}}}\right]=\int\mathcal{D}(\vecgrk{\Delta},\vecgrk{\rho})\tr_{\text{f}}\left[\e{-\beta \mathcal{\hat{H}}_{\text{BdG}}(\vecgrk{\Delta},\vecgrk{\rho})}\right]
\end{equation}
where the latter trace is over all fermionic degrees of freedom.
$\mathcal{H}_{\text{BdG}}(\vecgrk{\Delta},\vecgrk{\rho})$ is the
Bogoliubov-de Gennes (BdG) Hamiltonian with a given set of
$\vecgrk{\Delta}$ and $\vecgrk{\rho}$, where these vectors designate
the set of values of these parameters on all lattice sites
\begin{align}
&\hat{\mathcal{H}}_{\text{BdG}}=\sum_{i,\sigma}(\epsilon_{i}+\rho_{i})c^{\dagger}_{i\sigma}c_{i\sigma}
-\!\!\sum_{\langle i,j\rangle,\sigma}\!\left(t_{ij}c^{\dagger}_{i\sigma}c_{j\sigma}+t_{ij}^{*}c^{\dagger}_{j\sigma}c_{i\sigma}\right)\neweqnline
 &+\sum_{i}\left(\Delta_{i}c^{\dagger}_{i\uparrow}c^{\dagger}_{i\downarrow}+\bar{\Delta}_{i}c_{i\downarrow}c_{i\uparrow}\right)+\sum_{i}\frac{|\Delta_{i}|^{2}+\rho_{i}^2}{U_{i}}\punc{.}
 \label{eqn:BdGHamiltonian}
\end{align}

Given the explicit form of the diagonalizable BdG Hamiltonian, we can
calculate expectation values and correlation functions,
\begin{equation}
\tr\left[\hat{\rho}\ \!\hat{\mathcal{O}}\right]=\int\mathcal{D}(\vecgrk{\Delta},\vecgrk{\rho})\e{-\beta E_0}\sum_{n=1}^{N}\e{-\beta E_n}\left\langle n\left|\hat{\mathcal{O}}\right|n\right\rangle,
\label{eq:rhoO}
\end{equation}
where the sum is taken over all positive eigenvalues (quasi-particle
excitations) of the BdG Hamiltonian. Here $E_0$, $E_n$ and $|n\rangle$
are the ground-state energy, excitation energies and excitation wave
functions, respectively, for the BdG Hamiltonian, for the specific
configuration of $\vecgrk{\Delta}$ and $\vecgrk{\rho}$.  It is
straightforward to see that in this case, the saddle-point
approximation of the partition function gives rise to the mean-field
BdG equations (and then $\Delta_i$ indeed corresponds to
$|U_{i}|\langle c_{i\downarrow}c_{i\uparrow}\rangle$).
The calculation of the full integral, using the
(classical) Monte Carlo approach~\cite{Metropolis53}, includes also
the contributions of thermal fluctuations of the amplitude and phase
of the order parameter. In Appendix~\ref{MCsection} we detail how we
improve on contemporary methods to perform the Monte Carlo calculation
in $\mathcal{O}(N^{1.9}M^{2/3})$ time, where $N$ is the number of
sites and $M$ the order of a Chebyshev expansion.

\section{Applications}

We have derived a new expression for the current flow through a
superconductor, and demonstrated how to calculate the current in
mesoscopic systems.  Before applying it to understand and predict
novel phenomena, it is important to verify it across a variety of
exemplar systems, where one can compare against well-established
theories. As the main novelty of the approach is the inclusion of
thermal fluctuations, we pay particular attention to verifying the
formalism in two dimensions, especially looking for signatures of the
BKT transition driven by phase fluctuations. In
\secref{sec:LengthDependence} we probe the current through a clean
superconductor, and check that we can recover the BTK results for the
contact resistance. A key effect in such systems is Josephson
tunneling, so in \secref{sec:JosephsonJunction} we study the
temperature dependence of the resistance of a single
Josephson junction. Dephasing (by temperature averaging) and
decoherence (by electron-electron interactions) are studied in
\secref{sec:SelfEnergy}. In \secref{sec:KosterlitzThoulessTransition} we
examine the temperature dependence of the resistance on the two sides
of the BKT transition and compare to analytical results.
In \secref{sec:LittleParksEffect} we
introduce finite magnetic field (flux) and probe the Little-Parks
effect in the presence of disorder. Finally, in
\secref{sec:CurrentMaps} we demonstrate how plotting maps of the
current and potential across the system can illuminate the microscopic
processes at the superconductor-insulator transition. Throughout we
use an attractive interaction of $U=1.6t$ to describe the SC
region. To avoid the Van Hove singularity at
half filling~\cite{Scalettar89} we study systems at an average
$38.7\%$ filling, except for \secref{sec:SelfEnergy} where we focus on
wires with a low filling fraction of $20\%$. In the linear response
regime we impose a potential difference of
$eV=0.02t$. Systems were typically
two-dimensional, so a single lattice site thick, 12 lattice sites
wide, and 48 lattice sites long.

\subsection{Clean systems}\label{sec:LengthDependence}

\begin{figure}
 \includegraphics[width=0.95\columnwidth]{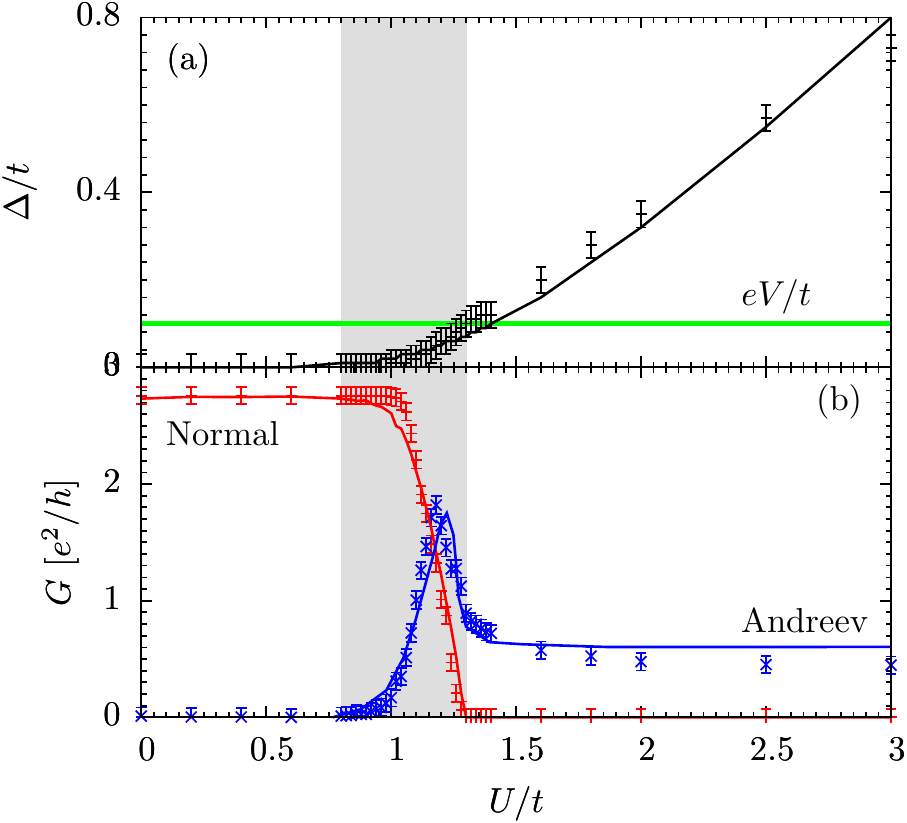}
 \caption{(Color online) The lower graph (b) shows the variation of
 conductance $\sigma$ with interaction strength. The normal current is shown in red
 and the supercurrent in blue. The numerical results are shown with
 error bars, and BTK theory by the solid line. The upper plot (a)
 shows the variation in the order parameter with interaction strength;
 numerics are shown with error bars and the mean-field theoretical
 prediction by the solid black line. The horizontal green line denotes
 the chemical potential difference across the sample. The gray shading denotes the range of
 interactions where $0<\Delta<eV$, so that both normal and Andreev
 processes contribute to the current.}
 \label{fig:BTKPlot}
\end{figure}	

At low temperatures, thermal fluctuations of the pair amplitude and
phase may be neglected.  The conductance, in this limit, through a
clean SC region has been calculated by Blonder, Tinkham and Klapwijk
(BTK) \cite{Blonder82}, and is solely due to the contact resistance at
the two interfaces. By assigning a tunneling strength $1/Z$ to the
barriers, BTK have shown that if the intermediate sample is in the
normal state, then the current is purely due to electrons tunneling
across the barrier, and the transmission coefficient is given by
$1/Z^{2}$~\cite{Blonder82}. On the other hand, if the sample is in the
SC state, then the SC gap, $\Delta$, inhibits electrons from directly
tunneling into it. Instead, these electrons Andreev tunnel accompanied
by a hole. For a large barrier $Z\gg1$ the transmission coefficient
becomes $\Delta^{2}/4Z^{4}(\Delta^{2}-E^{2})$~\cite{Blonder82},
where $E$ is the electron energy.  Electrons with an energy outside of
the gap can either tunnel alone with a corresponding normal
transmission coefficient
$(E+\sqrt{E^{2}-\Delta^{2}})/(2Z^{2}\sqrt{E^{2}-\Delta^{2}})$, or
Andreev tunnel with accompanying hole, and have a transmission
coefficient of $\Delta^{2}/4Z^{4}(E^{2}-\Delta^{2})$. We first
compare the results of our numerical calculations to these BTK
formulae, and then demonstrate that for the simple case of a single SC
site, the BTK results can be derived analytically from our current
formula.

To verify that the model recovers the correct behavior at the
tunneling barrier we focus on the weak coupling limit. In this limit,
once a Cooper pair tunnels through the first barrier, it has an equal
probability of continuing to either the left or the right lead, an
consequently the current through the double barrier will be half that
of a single barrier~\cite{Lambert91}.  For a long enough system the
finite bias and temperature smear any Fabry-Perot type
interference. To study the effect of the changing order parameter
$\Delta$, we focus on a $39\%$ filled system at ``zero'' temperature
(without quantum fluctuations), where $\Delta$ is indeed equal to the
pair correlation $|U|\langle c_{i\downarrow}c_{i\uparrow}\rangle$,
vary the interaction strength $U$, and monitor the various components
of the tunneling current. For a pristine system with $W=0$, all of the
resistance stems from the two tunneling barriers, and we verified that
the current flow was independent of the length of the SC region.  A
relatively large potential bias of $eV=0.1t$ was applied across the
leads. This allows us to explore all tunneling processes, either for
$\Delta<eV$ or $\Delta>eV$ by changing the
interaction parameter $U$ and as a result $\Delta$, see
\figref{fig:BTKPlot}(a). Our results for the current are depicted in
\figref{fig:BTKPlot}(b), and has $Z\approx1.4$. At $U=0$ the current
is entirely normal. As shown in \figref{fig:BTKPlot}(a), with
increasing $U$ the SC gap grows giving rise to a resonance in the Andreev
current when $E=\Delta$. At the same time the normal current falls as fewer electrons can
be directly injected outside of the SC gap. As the interaction
strength is increased further, so that the SC gap exceeds the chemical
potential difference, resonant electrons are no longer injected into
the divergent density of states at the SC gap, and the Andreev current
falls. In agreement with the BTK calculation, at large $\Delta$ the
Andreev current adopts its final value, $1/4$ of the normal $U=0$
conductance and no normal current flows. The agreement with the BTK
prediction verifies that the current formula
\eqnref{eqn:PerturbativeEquationForCurrent} contains the
correct tunneling behavior.

We now turn to derive the BTK results from our formalism analytically,
which can be done straightforwardly in the weak coupling limit $Y\ll1$ when
the SC region consists of a single site, and we take the leads as
having a parabolic dispersion. In the linear response regime
where a potential $V$ is put across the sample such that
injected electrons are entirely within the SC gap, we find that the
normal current is zero and we recover the analytic result for the
Andreev current,
$J=e^{2}V\Delta^{2}/8hZ^{4}(\Delta^{2}-\mu^{2})$, where $Z=\sqrt{\mu/\pi\nu}/Y$,
$\mu$ is the chemical potential, and $\nu$ is the density of states at the Fermi surface. If the
sample is normal we find that there is no Andreev current, and the
normal current is $J=e^{2}V/2hZ^{2}$. These results are what would
be expected from the BTK formalism, and coupled with the numerical
results confirm that the formalism properly treats tunneling between
the leads and the SC sample.

\subsection{Josephson junction}\label{sec:JosephsonJunction}

\begin{figure}
 \includegraphics[width=0.95\columnwidth]{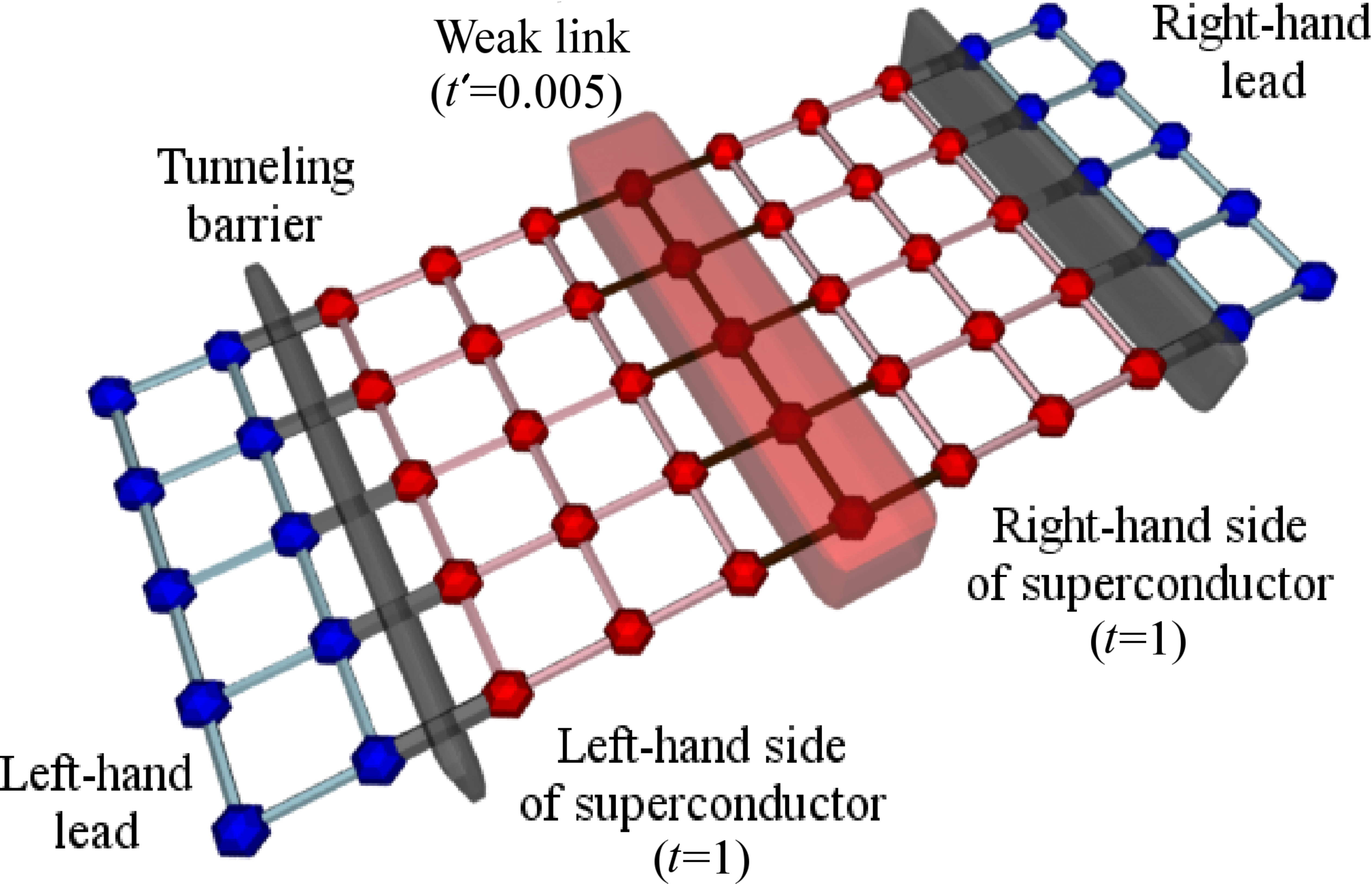}
 \vspace{5 mm}
 \includegraphics[width=0.95\columnwidth]{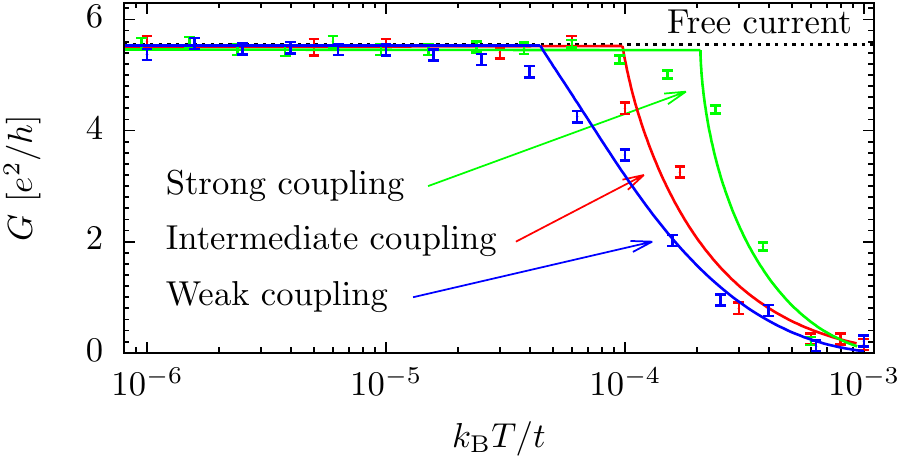}
 \caption{(Color online) \textit{Upper}: The setup to model the
 Josephson junction.  Traversing the center of the SC region is a
 Josephson junction (opaque cuboid). The junction is modeled by the
 reduction of the matrix hopping elements to $t'=0.005$ (brown
 interconnects) compared to $t=1$ in the superconductor. The two
 metallic leads are shown in blue, and the lead-superconductor
 tunneling barrier by the gray cuboids.  \textit{Lower}: The variation
 of conductance with temperature for the Josephson junction. Results
 of the numerical computation (points) and of the theoretical model
 (solid lines) are shown for a weak (green), intermediate (red), and
 strong (blue) coupling between the two superconductors.}
 \label{fig:JosephsonPlot}
\end{figure}

Another simple example that we wish to explore is a single Josephson
junction, which will be modeled in the negative-$U$ Hubbard model by
an intermediate region consisting of two clean superconductors,
between which we insert a weak link where the nearest-neighbor hopping
element $t'$ is small ($t'<<t$), see
\figref{fig:JosephsonPlot}. Studying this system will allow us to
probe how a phase difference across a barrier can affect the current
flow through it.  We again adopt a $39\%$ filled band with
no disorder. We first set $t'=0$ to disconnect the left and right-hand
sides, and numerically evaluate the current through the central
region. This is by no means trivial. The current formula, through the
anomalous Green function, allows an absorption of a pair from the
incoming lead into the condensate on one side of the barrier, and an
emission of another pair into the outgoing lead. This current,
however, will depend on the phase difference between the
SC order parameter on the two sides of the barrier. For
$t'=0$, i.e.  an infinite barrier, the phases of the left and
right-hand order parameter are uncorrelated, and thus all phase
differences are degenerate in energy. Therefore, the current vanishes, but only 
after averaging over all states, which is done automatically in our
numerical procedure.  In the other limit, when the hopping matrix
elements are the same as the hopping through the rest of the
superconductor, $t'=t$, the phase of the superconductor is locked so
we see the standard free SC current flow.

We now model the situation with a moderately sized central barrier.
This splits the superconductor in two, but crucially a Josephson
supercurrent flows between the two sides, thus allowing the current to
flow with no additional resistance. The current
$J(T)=J_{\text{J}}(T)\cos(\phi_{\text{L}}-\phi_{\text{R}})$ is maintained by
the phase difference $\phi_{\text{L}}-\phi_{\text{R}}$ between the
left and right-hand superconductors, and the maximum value of the
dissipationless current is the critical Josephson current
$J_{\text{J}}(T)=(\pi|\Delta|/2eR_{\text{n}})\tanh(|\Delta|/2k_{\text{B}}T)$~\cite{Ambegaokar63},
where $R_{\text{n}}$ is the resistance of the central barrier when
the system is in the normal state.

In order to study the thermally driven disruption of the Josephson
current numerically, it is vital that this breakdown occurs before the
BKT transition occurs, which as we show in
\secref{sec:KosterlitzThoulessTransition}, by itself reduces the current flow
through the system. We thus use a small hopping element for the
tunneling barrier of $t'=0.005t$, which has a large $\rho_{\text{n}}$ and
therefore small Josephson current $J_{\text{J}}$.  In
\figref{fig:JosephsonPlot} we show the current as a function of
temperature, in the presence of the weak link.  When there is no
voltage drop across the Josephson junction, the current $J_{\text{M}}$
that flows through it is given by $J_{\text{M}}=V/R$, where
$R$ is the contact resistance to the normal leads. This current is
maintained as long as the critical Josephson current $J_{\text{J}}$ is larger than
$J_{\text{M}}$. As temperature is increased thermal fluctuations will
weaken the phase lock between the two superconducting regions, and the
critical current is reduced. When $J_{\text{J}}$ is reduced below
$J_{\text{M}}$, a finite voltage develops across the Josephson
junction. This drives the phase difference across the junction to
increase with time, which in turn leads to an oscillating
current. This current has a non-zero time-average
\cite{BCS}, leading to a total resistance
$R_{\text{JJ}}(T)=R/(1-\sqrt{1-\lambda^{2}})$, where
$\lambda=J_{\text{J}}/J_{\text{M}}<1$~\cite{BCS}. This time
averaged current is exactly the quantity calculated in our Monte Carlo
procedure. \figref{fig:JosephsonPlot} depicts a comparison between this simple model
and our full numerical calculation, with reasonable agreement.

The critical current can be modified by varying the resistance of the
central barrier, $R_{\text{n}}$. The intermediate case has
$R_{\text{n}}=0.075h/e^{2}$, the stronger coupling with
$R_{\text{n}}=0.06h/e^{2}$ is obtained by lowering the barrier to
$t'=0.01t$, and the weaker coupling with
$R_{\text{n}}=0.085h/e^{2}$ by widening the original barrier
($t'=0.005t$) to four lattice sites. This wider barrier weakens the coupling
between the superconductors so the Josephson resistance emerges at a
lower temperature. A lower barrier strengthens the coupling so raises the temperature
required for the emergence of resistance. Both these regimes are
consistent with our simple model. We also verified that at very strong coupling
where the temperature required for the breakdown of phase coherence becomes
of the order of the BKT transition temperature our simple model for
the current flow no longer captures the full physics of the system. This
study validates that our formalism can correctly model the presence of
the Josephson supercurrent across the weak link introduced into the
superconductor, and can therefore be used to model mesoscopic systems
that contain multiple SC grains.

\subsection{Decoherence and dephasing}\label{sec:SelfEnergy}

\begin{figure}
 \includegraphics[width=0.95\columnwidth]{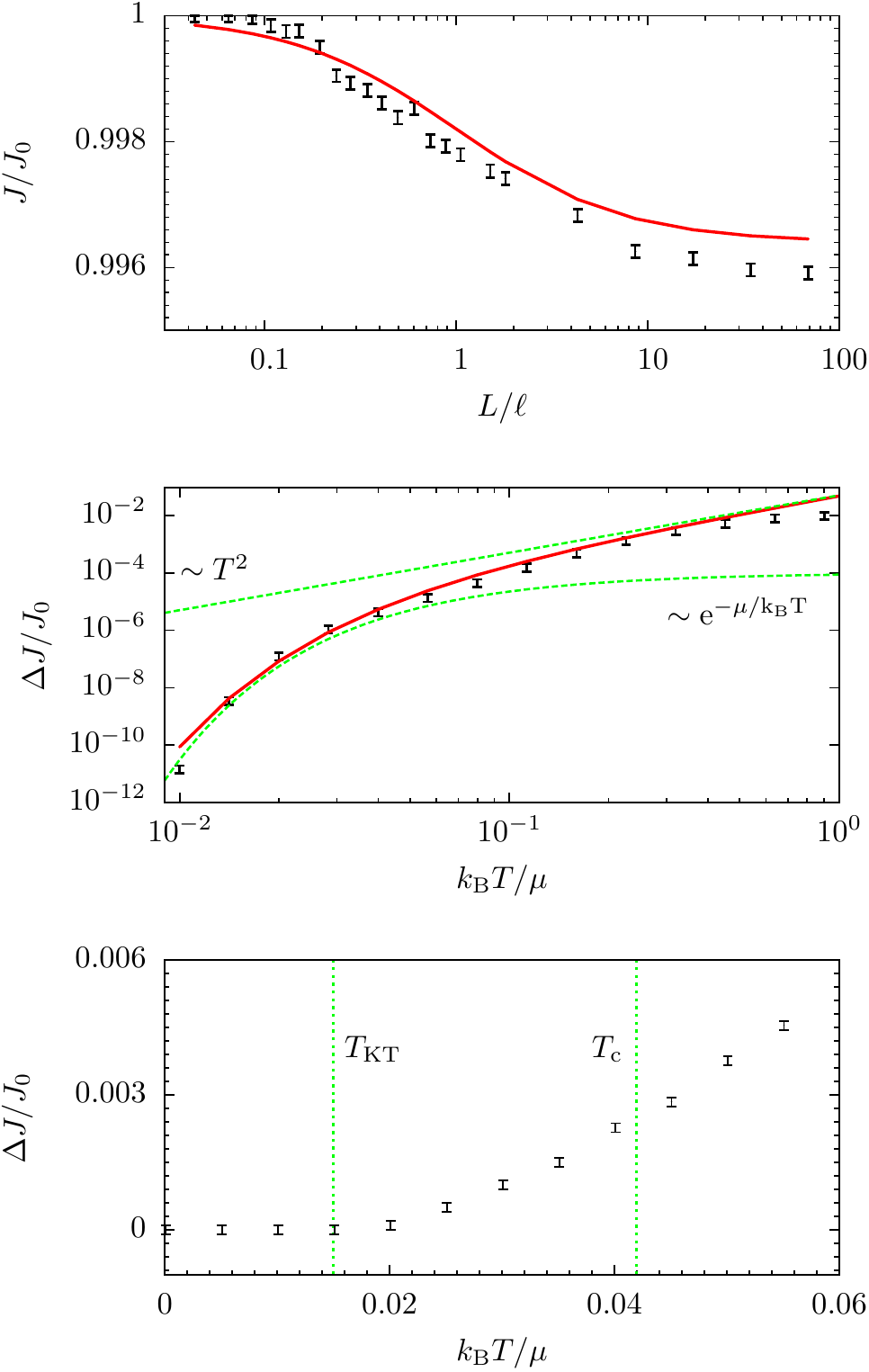}
 \caption{(Color online) (a) The relative fall in current in a
 one-dimensional sample due to the introduction of self energy at
 $k_{\text{B}}T=0.1\mu$. The black points show the numerical results, and the red
 line highlights the expected theoretical variation with length
 \cite{Micklitz10}, where $J_{0}$ is the current that flows when
 impeded solely by the contact resistance.
 (b) The relative change ($\Delta J\equiv J_0-J$)
 in current as a function of temperature in a one-dimensional normal
 phase sample. The black points show the numerical results, and the
 red the expected model variation. The two green dashed lines show the
 $\exp(-\mu/k_{\text{B}}T)$ and $T^{2}$ behavior. (c) The changing current in the
 presence of a SC phase in a two-dimensional sample. The vertical
 green dashed lines show the BKT and normal phase transitions.}
 \label{fig:SelfEnergyPlot}
\end{figure}

The issue of decoherence and dephasing plays a significant role in
transport at low temperatures. Here we define decoherence as the
many-body phenomenon that leads to the loss of coherence via
interactions among the electrons or interactions with the
environment. On the other hand, dephasing can occur in a
non-interacting system, and emerges from the fact that due to the
finite temperature, electrons possess a range of energies, of the
order of $k_{\text{B}}T$.  Electrons of different energies acquire
different phases along their respective trajectories, and if these
phases differ by $2\pi$ or more when their energy changes by
$k_{\text{B}}T$, then interference phenomena will average out to zero.

\emph{Decoherence}: The effects of decoherence due to
electron-electron interactions are more profound in one-dimensional
wires in the normal phase.  Since the original Hubbard model employed
in this calculation (\eqnref{eq:Hubbard}) is an interacting model, one
expects decoherence to arise naturally from the calculation. However,
though the original formula for the current is exact, the
approximation employed above -- \eqnref{eq:rhoO} -- does not include
quantum fluctuations. This means that it neglects the imaginary
component of the self energy which corresponds to damping due to
interactions, and the resulting decoherence. In order to test the
effects of such decoherence due to many-body interactions, we
introduce into the normal Green function for momentum $k$ (as here we study a wire in
the normal phase), by hand, the self energy
\begin{equation}
 \lim_{\delta\to0}\frac{U^{2}}{2\pi^{4}}\sum_{\vec{p},\vec{q}}\frac{n(\xi_{\vec{p}})[1-n(\xi_{\vec{p}-\vec{q}})][1-n(\xi_{\vec{k}+\vec{q}})]}
 {\omega+\xi_{\vec{p}}-\xi_{\vec{p}-\vec{q}}-\xi_{\vec{k}+\vec{q}}-\cmplxi\delta}\punc{,}
\end{equation}
which is the lowest order contribution to the single-particle
self-energy, and where $\xi_{\vec{p}}$ are the momentum energy
eigenstates of the Hamiltonian.

A similar approach \cite{Micklitz10} has been applied to interacting
electrons in a continuous one-dimensional system with repulsive
contact interactions (the second order contribution to the self-energy
does not depend on the sign of the interaction). In this case, it has
been shown, for wires with parabolic dispersion and
chemical potential $\mu$, that this damping
leads to a change in the distribution function and reduction in the
conductivity by a factor of
$1-\pi^{2}(k_{\text{B}}T/\mu)^{2}L/12[L+\ell\exp(\mu/k_{\text{B}}T)]$~\cite{Micklitz10}, where
the wire length, $L$, is long enough that the smearing of the Fermi
surface due to scattering (that occurs over the relaxation length
scale $\ell$~\cite{Micklitz10}) outweighs that due to temperature. For
sufficiently long wires $L\gg\ell\exp(\mu/k_{\text{B}}T)$ the reduction in
conductivity becomes length independent $1-\pi^{2}(k_{\text{B}}T/\mu)^{2}/12$.

In order to be able to compare to this theory (which relies on the
parabolic dispersion), we focus on a system with a low filling
fraction of $20\%$, near the bottom of the band, and set the disorder
to $W=0.1t$. We employ the perturbative expression for the
current, \eqnref{eqn:PerturbativeEquationForNormalCurrent},
to give us access to long wires with $L\gg\ell$.
In the upper panel of \figref{fig:SelfEnergyPlot} we show
the fall in current, as a function of length, due to the inclusion of
self energy at $k_{\text{B}}T=0.07\mu$, here $\ell\approx27a$.
The overall change of $\sim0.5\%$ is small
due to the Pauli blocking of scattering processes near the Fermi
energy.  We see reasonable agreement with the model over a range of
length scales.  The middle panel of \figref{fig:SelfEnergyPlot}
depicts the change in current with temperature for a system of a fixed
length.  We highlight the agreement to the expected variation in the
fall in conductance with temperature \cite{Micklitz10}.  At low
temperatures ($k_{\text{B}}T\ll\mu$) the damping is severely Pauli blocked so the
characteristic damping length-scale exceeds the system length and the
current correction $1-\pi^{2}(k_{\text{B}}T/\mu)^{2}\exp(-\mu/k_{\text{B}}T)L/12\ell$ is
exponentially suppressed. As temperature increases the Fermi liquid
$T^{2}$ behavior starts to dominate the correction to the
current. At high, usually unphysical temperatures ($k_{\text{B}}T\gg \mu$),
numerics see a smaller current shift than predicted by theory as the
details of the specific Hubbard band dispersion versus the parabolic
dispersion in which the model was developed become important.

In \figref{fig:SelfEnergyPlot}(c) we examine the effect of a SC phase
on decoherence.  At low temperature the presence of the SC gap
suppresses many-body scattering processes. However, when temperature
is raised above the BKT phase transition, scattering events are
possible, though have a smaller impact on the current than in the
normal phase, due to the still finite local pair correlations.  Above
the mean-field BCS phase transition the current follows the expected
parabolic profile as in the normal phase. Thus we have demonstrated
that while quantum fluctuations, as they affect decoherence, can be
taken into account in our formalism, their effect on the current, for
the range of parameters studies here, is usually small at
$\lesssim1\%$. We are thus justified in neglecting them in this study.

\begin{figure}
 \includegraphics[width=0.95\columnwidth]{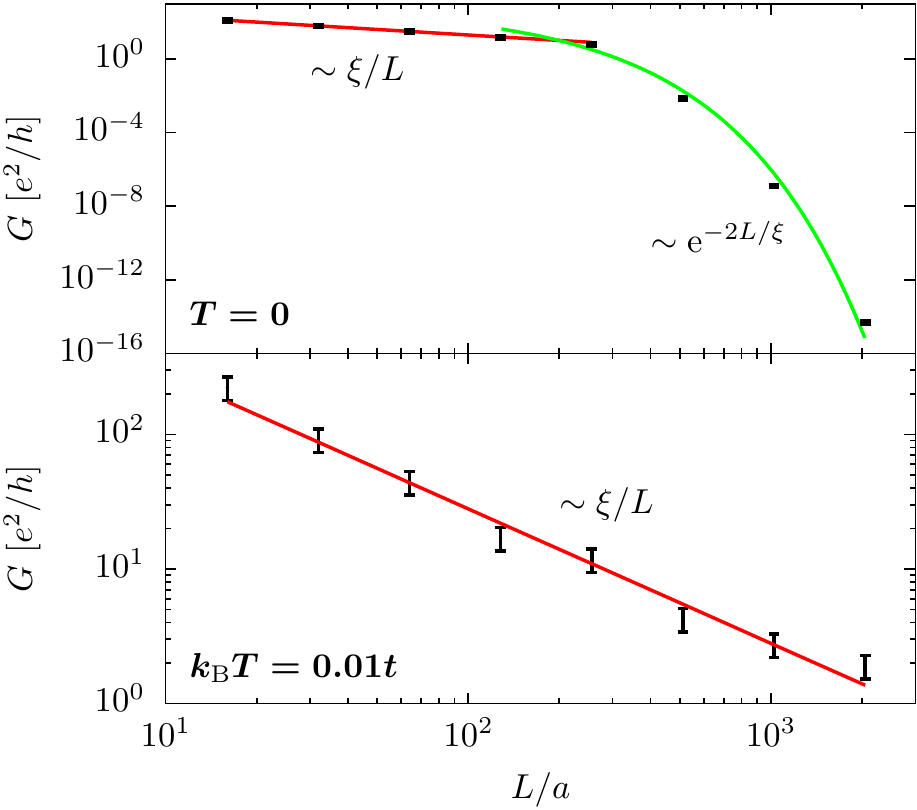}
 \caption{(Color online) The fall in conductance with length in a
 non-interacting one-dimensional wire with disorder $W=0.2t$ at two
 different temperatures. The red trend
 lines show a linear drop off in conductance with length, and green an
 exponential decay.}
 \label{fig:NonIntRes}
\end{figure}

\emph{Dephasing}: Having observed decoherence in the sample we now turn to study dephasing due to thermal
averaging. To verify that our formalism captures this important
phenomenon, we study the length dependence of the conductance in
non-interacting systems.  We first verified, for a non-interacting
clean system, that the net macroscopic current increases by $2e^{2}/h$
for each new conduction channel introduced (not shown), independent of
length. Setting the amplitude of the disorder to $W=0.2t$ and working
at $39\%$ filling, in \figref{fig:NonIntRes} we show the fall in
conductance with length at two different temperatures. At $T=0$ there
is an initial linear fall in conductance over length scales smaller
than the localization length $\xi\approx93a$, and an exponential fall at greater
lengths. This is in accordance with the expectations of Anderson
localization~\cite{Ong01} -- for length scales below the localization
length, the conductance changes as a power law of the length, while it
decays exponentially when the length becomes larger than the
localization length. At $k_{\text{B}}T=0.01t$, on the other hand, dephasing causes
different parts of the system to be incoherent with respect to the
others, causing the conductance to fall linearly with inverse length,
as expected from a classical system.  This observation confirms that
the formalism naturally incorporates the physics of dephasing in
disordered systems.

\subsection{Variation of resistance with temperature}\label{sec:KosterlitzThoulessTransition}

\begin{figure*}
 \includegraphics[width=0.95\linewidth]{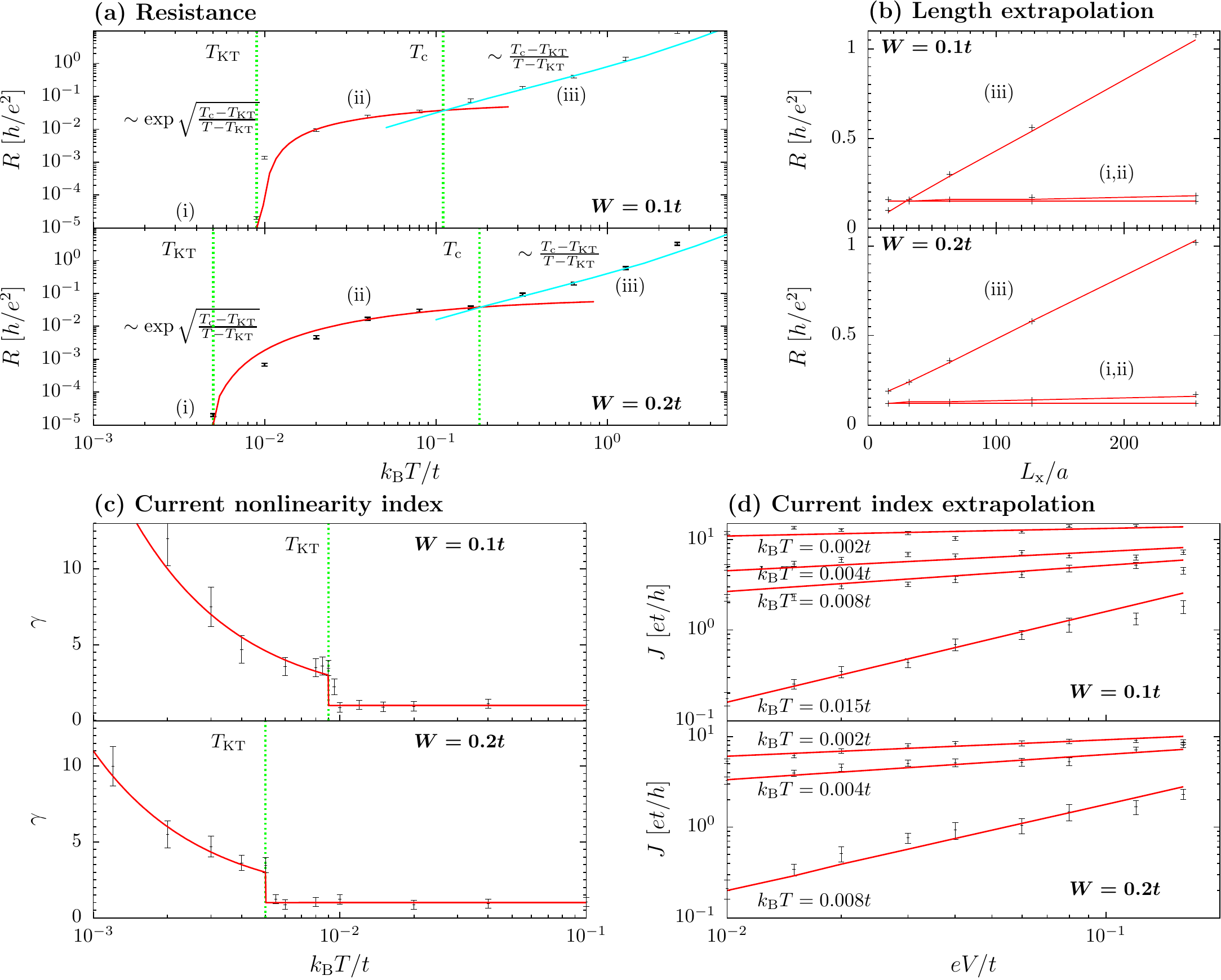}
 \caption{(Color online) (a) The
 variation of resistance with temperature for two different values of
 disorder calculated numerically (points). The red solid line shows the theoretical
 low temperature behavior, and the blue solid line the theoretical high
 temperature behavior. The dashed vertical green lines show the BKT
 $T_{\text{KT}}$ and mean-field $T_{\text{c}}$ temperatures.  (b) The
 numerical results (black points) and the deduced linear length
 dependence (red line) of the resistance for three different points
 (i) $T<T_{\text{KT}}$, (ii) $T_{\text{KT}}<T<T_{\text{c}}$, and (iii)
 $T>T_{\text{c}}$.  (c) The current nonlinearity index $\gamma$ in
 $V\propto J^{\gamma}$ for two disorder levels, as a function of
 temperature. The black points are the numerical results and the red
 lines are from theory. The vertical green dashed line highlights the
 BKT transition temperature. (d) Examples of the numerical
 measurements of dimensionless current against voltage leading to the
 values of $\gamma$ shown in the left-hand plot. The best-fit lines
 employed are shown in red.}  \label{fig:KTPlot}
\end{figure*}

We have now verified that our formalism captures the basic phenomena
of contact resistance in \secref{sec:LengthDependence}, Josephson
coupling in \secref{sec:JosephsonJunction}, and dephasing and
decoherence in \secref{sec:SelfEnergy}. With these key tests complete,
we are now ideally poised to study further effects within the
superconductor, starting with the temperature dependence of the
conductivity and its relation to the BKT transition. With increasing
temperature a two-dimensional superconductor undergoes a BKT
transition~\cite{{Kosterlitz73}} characterized by the emergence of
vortices across the system, leading to the loss of global phase
coherence. At a higher (``mean-field'') temperature, the SC order is
completely suppressed and the system loses the SC correlation even
locally. To study how this transition is reflected in the current flow
we performed numerical simulations on a two-dimensional $39\%$ filled SC
system at several different temperatures. Simulations were performed
for two different levels of disorder, $W=0.1t$ and $W=0.2t$, to
determine how the transition and current flow are modified by the
normal-state resistance, and extrapolated over length to remove the
effects of the contact resistance (\figref{fig:KTPlot}(b)).

Even at temperatures below the BKT transition, vortices can be
nucleated from the edge of the sample and traverse the system, driven
by the Magnus force due to the finite current. This produces
dissipation at any non-zero temperature and current $J$, according to
the non-linear potential $V\propto
J^{1+2T_{\text{KT}}/T}$~\cite{Halperin79}. In \figref{fig:KTPlot}(a)
we see that below the BKT temperature the linear resistance, that is
$\lim_{V\to0}V/J$, is zero. The plots \figref{fig:KTPlot}(d) show
several simulations that were performed for different imposed
potential differences $V$ across the sample, which allowed us to
extract the index $\gamma$ of the conductance relation $V\propto
J^{\gamma}$. In \figref{fig:KTPlot}(c) we show that the conductance
relation approximately follows the expected theoretical behavior with
$\gamma=1+2T_{\text{KT}}/T$.

At temperatures above the BKT transition vortices and anti-vortices
can easily unbind, though they may be partially pinned by
disorder. The finite conductance $G$ of a sample in this case has been shown
by Halperin and Nelson \cite{Halperin79} to be given by
\begin{equation}
 G=0.37G_{\text{n}}(\ksi_+/\ksi_{\text{c}})^2\punc{,}
\end{equation}
where $G_{\text{n}}$ is the normal state conductance,
$\ksi_{\text{c}}$ is the SC coherence length, and $\ksi_+$ is the SC
order correlation length, which diverges at $T_{\text{KT}}$. The
critical behavior at temperatures near the BKT transition $T\gtrsim
T_{\text{KT}}$ leads to the conductance
\begin{equation}
 G=0.37G_{\text{n}}b^{-1}\exp[\sqrt{b(T_{\text{c}}-T_{\text{KT}})/(T-T_{\text{KT}})}]\punc{,}
 \label{sigmaKT}
\end{equation}
where $b$ is a number of order unity. At temperatures higher than the
(renormalized) mean-field critical temperature $T_{\text{c}}$, the
conductance is given by the Aslamasov-Larkin
theory~\cite{Aslamasov68,Halperin79}
\begin{equation}
 G=0.37G_{\text{n}}(T_{\text{c}}-T_{\text{KT}})/(T-T_{\text{KT}})\punc{.}
 \label{sigmaAL}
\end{equation}
Finally, we can also estimate the crossover between these two regimes
by noting that the difference between the Kosterlitz-Thouless and
mean-field transition temperatures critical regime is given
by~\cite{Halperin79}
\begin{equation}
 T_{\text{c}}-T_{\text{KT}}\approx0.17e^{2}T_{\text{c}}/\hbar\sigma_{\text{n}}\punc{.}
 \label{eqn:TDiff}
\end{equation}
The difference between these two temperatures therefore widens with
falling normal state conductance.

In
\figref{fig:KTPlot}(a) we depict the variation of resistance
with temperature above the BKT transition, showing the two types of
dependence on temperature as is expected by theory. We also note that
the rising disorder increases the normal state resistance
$\sigma_{\text{n}}$, and also broadens the difference between the
Kosterlitz-Thouless and mean-field transition temperatures, which
agrees with \eqnref{eqn:TDiff} within $~20\%$. The high temperature
Aslamasov-Larkin expression for the conductance persists well above
the the mean-field critical temperature, where the SC state has been
totally suppressed. Finally, when the temperature is of the same order
as the bandwidth, $k_BT\sim t$, the resistance in \figref{fig:KTPlot}(a)
increases superlinearly as the Fermi distribution becomes smeared
across the whole band structure.

\begin{figure}
 \includegraphics[width=0.95\columnwidth]{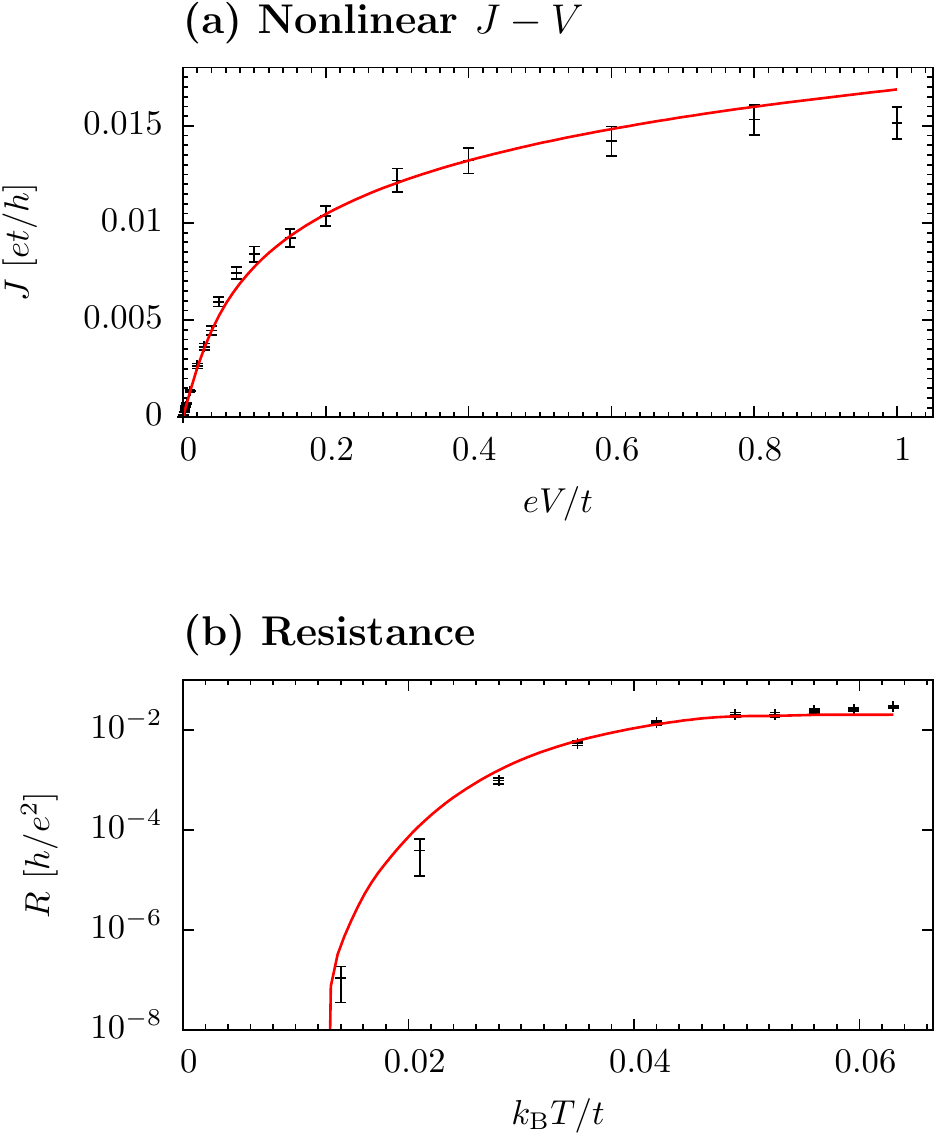}
 \caption{(Color online) (a) The nonlinear $J-V$ characteristic of a
 one-dimensional superconductor at fixed temperature. The numerical
 points are shown with black error bars and the
 Langer-Ambegaokar-McCumber-Halperin model~\cite{Langer67,McCumber70}
 by the red line. (b) The variation of resistance with temperature for
 fixed bias. The numerical points are shown with black error bars and
 the Langer-Ambegaokar-McCumber-Halperin model by the red line.}
 \label{fig:1DNonLinearIV}
\end{figure}

Having studied the nonlinear $J-V$ characteristic in two dimensions we
now turn to look at the one dimensional system. Here thermal
fluctuations can drive the formation of phase slips at any temperature
and so this system has the $J-V$ characteristic
$V=J_{0}R\sinh(J/J_{0})$~\cite{Langer67,McCumber70,Altomare05}, where
$J_{0}=4ek_{\text{B}}T/h$ and
$R=(h/4e^{2})\times(\hbar\Omega/k_{\text{B}}T)\exp(-\Delta
F/k_{\text{B}}T)$ is the resistance with attempt frequency $\hbar\Omega\approx3.1t$
and energy barrier $\Delta F\approx3.7t$. In \figref{fig:1DNonLinearIV}(a) we
show the consistency of the numerical model both for the nonlinear
$J-V$ characteristic, and in \figref{fig:1DNonLinearIV}(b) the
variation with temperature. The strong accord between analytics and
numerics in both one and two dimensions gives us confidence that the
formalism can be applied to study and explore less well understood
mesoscopic superconducting systems.

\subsection{Little-Parks effect}\label{sec:LittleParksEffect}

\begin{figure}
 \includegraphics[width=0.95\columnwidth]{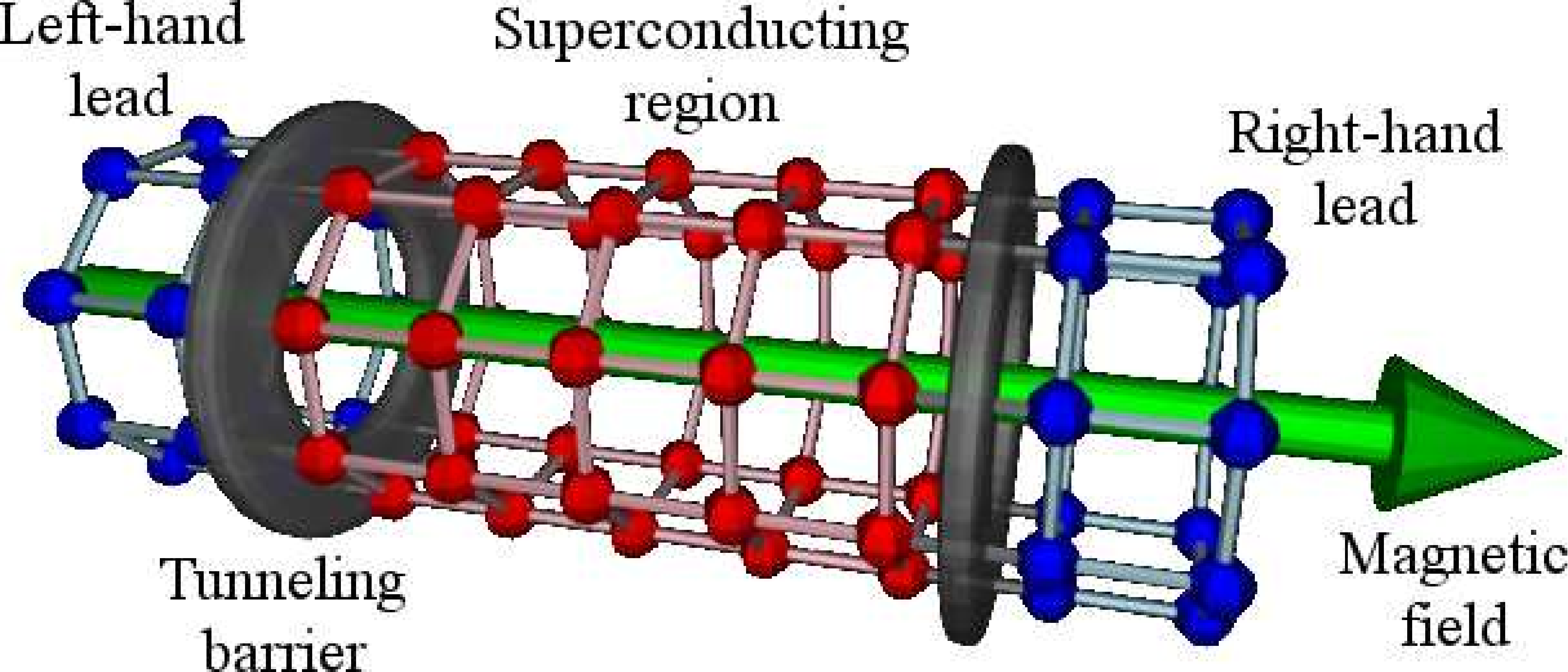}
 \vspace{10 mm}
 \includegraphics[width=0.95\columnwidth]{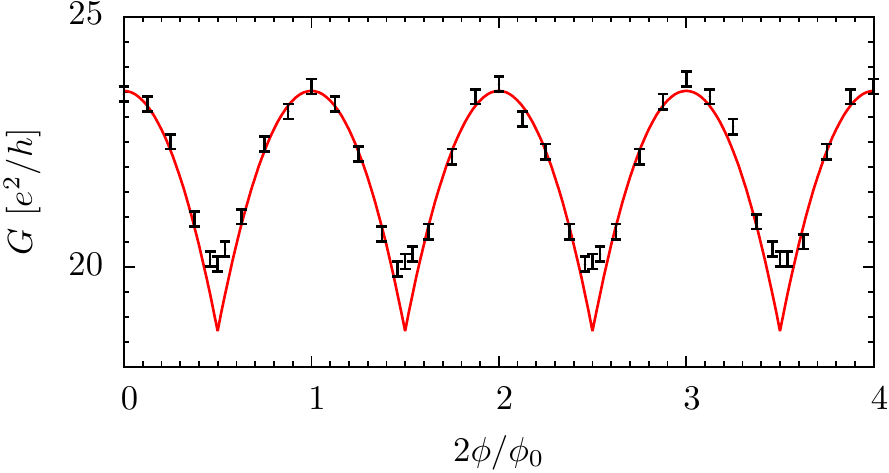}
 \caption{(Color online) \textit{Upper}: A schematic of the
 cylindrical wire within the negative-$U$ Hubbard model. The left and
 right-hand metallic leads are shown in blue, from which electrons can
 tunnel through the gray toroids into the central SC region which is
 shown in red. The magnetic flux threading the cylinder is shown in
 green.  \textit{Lower}: The variation of current with longitudinal
 magnetic flux at $T=T_{\text{c}}$. The computational points with
 error bars are shown in black, and the Little-Parks model best fit is
 shown by the red dashed line.}  \label{fig:LittleParksPlot}
\end{figure}

Varying an applied magnetic field has long been an important
experimental probe of the properties of a superconductor. It is
therefore imperative to verify that the current formula developed
here, coupled with the Hubbard model for the superconductor, is able
to accurately model the effects of an applied magnetic field. In the
Hubbard model the effects of the magnetic field are incorporated, via
the Peierls substitution, into the phases of the hopping elements,
$t_{ij}\rightarrow t_{ij}\e{2\pi\cmplxi\phi_{ij}/\phi_0}$ where
$\phi_0=hc/e$ is the quantum flux, and the phases $\phi_{ij}$ are
defined such that their integral over a closed trajectory is equal to
the magnetic flux threading the surface spanned by the trajectory.

In order to check whether this procedure captures the effect of an
orbital magnetic field, we apply it to a hollow cylindrical
superconductor, of radius $r$, such as that shown in \figref{fig:LittleParksPlot},
threaded by magnetic flux. As demonstrated by Little and
Parks~\cite{Little62}, the flux suppresses superconductivity and the
transition temperature falls periodically with the flux. This is often
probed by measuring the falling conductance of the cylinder near to
the transition temperature~\cite{Little62,Liu01}.

We apply our formalism to  the cylindrical thin-walled superconductor shown in
\figref{fig:LittleParksPlot} at $39\%$ filling and no disorder,
and apply an external magnetic flux $\phi$ along the axis of the
cylinder. The additional phase shift to the
hopping matrix elements around the cylinder circumference
causes the energy of electrons in the cylinder of radius $r$ to
increase with trapped flux $\phi$ as
$\hbar^{2}(n+2\phi/\phi_{0})^{2}/2mr^{2}$, where the integer $n$ is
chosen to minimize the energy. This results in a periodic parabolic
variation of the electron energy with flux and thus a parabolic
periodic oscillation in the SC transition temperature
$\Delta
T_{\text{c}}=\hbar^{2}(n+2\phi/\phi_{0})^{2}/16mr^{2}$~\cite{Little62}. Therefore,
for a cylinder held just below its superconducting transition
temperature, with increasing flux the superconducting state is
disrupted periodically and the resistance varies with flux, as a
series of parabolas, with minima in the conductance at every half
flux quantum $\phi=n\phi_{0}/2$.  This has indeed been observed
experimentally~\cite{Little62}.

In \figref{fig:LittleParksPlot} we take a cylinder held near to its SC
transition temperature and numerically evaluate the conductance as a
function of the magnetic flux. The reasonable agreement with theory
demonstrates that the formalism correctly picks up the effects of an
applied magnetic field. The deviation from the parabolic predictions
of mean-field theory at every half flux quantum is due to thermal
fluctuations, and will elaborated upon in a later publication.

\subsection{Current distribution maps}\label{sec:CurrentMaps}

\begin{figure}
 \includegraphics[width=0.95\columnwidth]{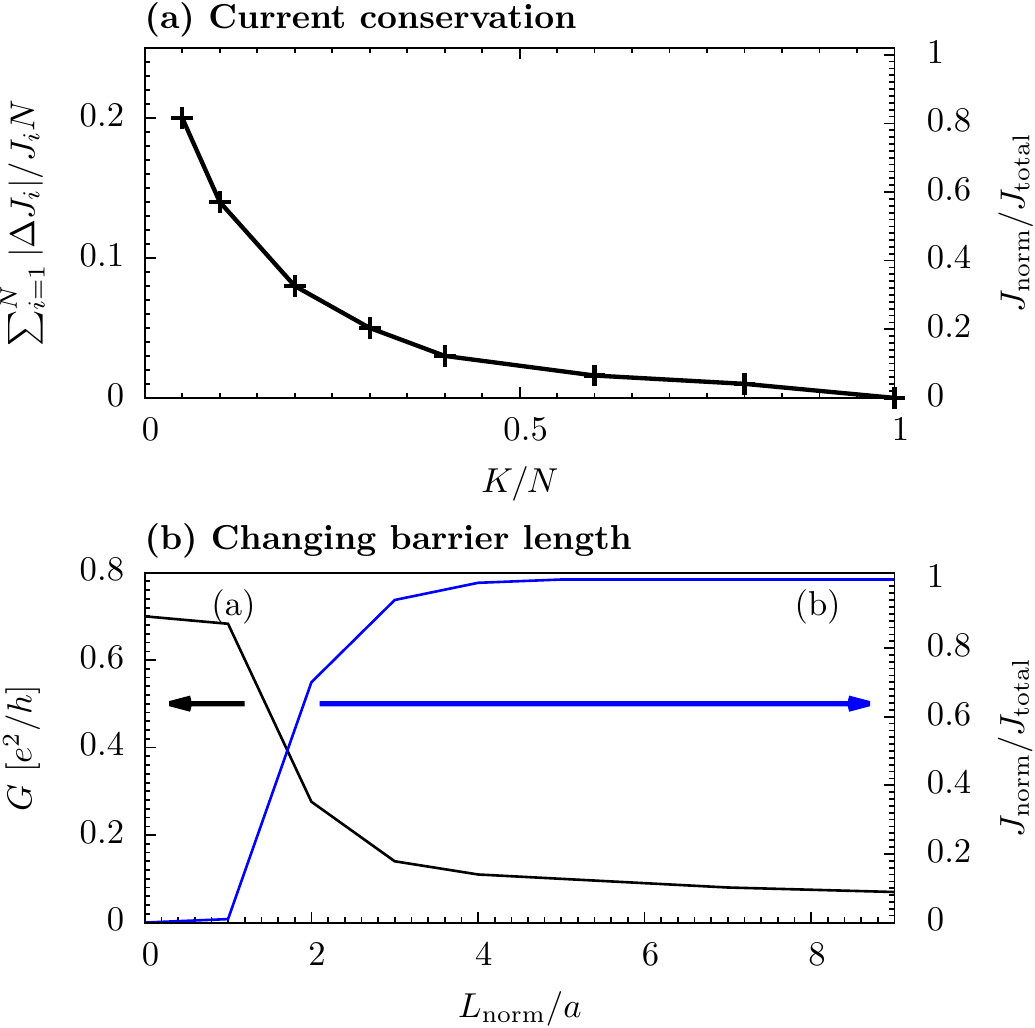}
 \caption{(Color online) (a) The average fractional error in
 conservation of current $\sum_{i=1}^{N}|\Delta J_{i}|/J_{i}N$ on each
 site against the fraction of total states $K/N$ included in the
 calculation of the current. (b) The changing conductance (black line)
 with width $L_{\text{norm}}$ of central normal region, the right axis
 shows the normal fraction $J_{\text{norm}}/J_{\text{total}}$ of the
 total current (blue line) flowing through the central region.}
 \label{fig:CurrentConservation}
\end{figure}

\begin{figure}
 \begin{flushleft} (a) Current map for a short barrier $L_{\text{norm}}=a$
 \end{flushleft}
 \includegraphics[width=0.55\columnwidth]{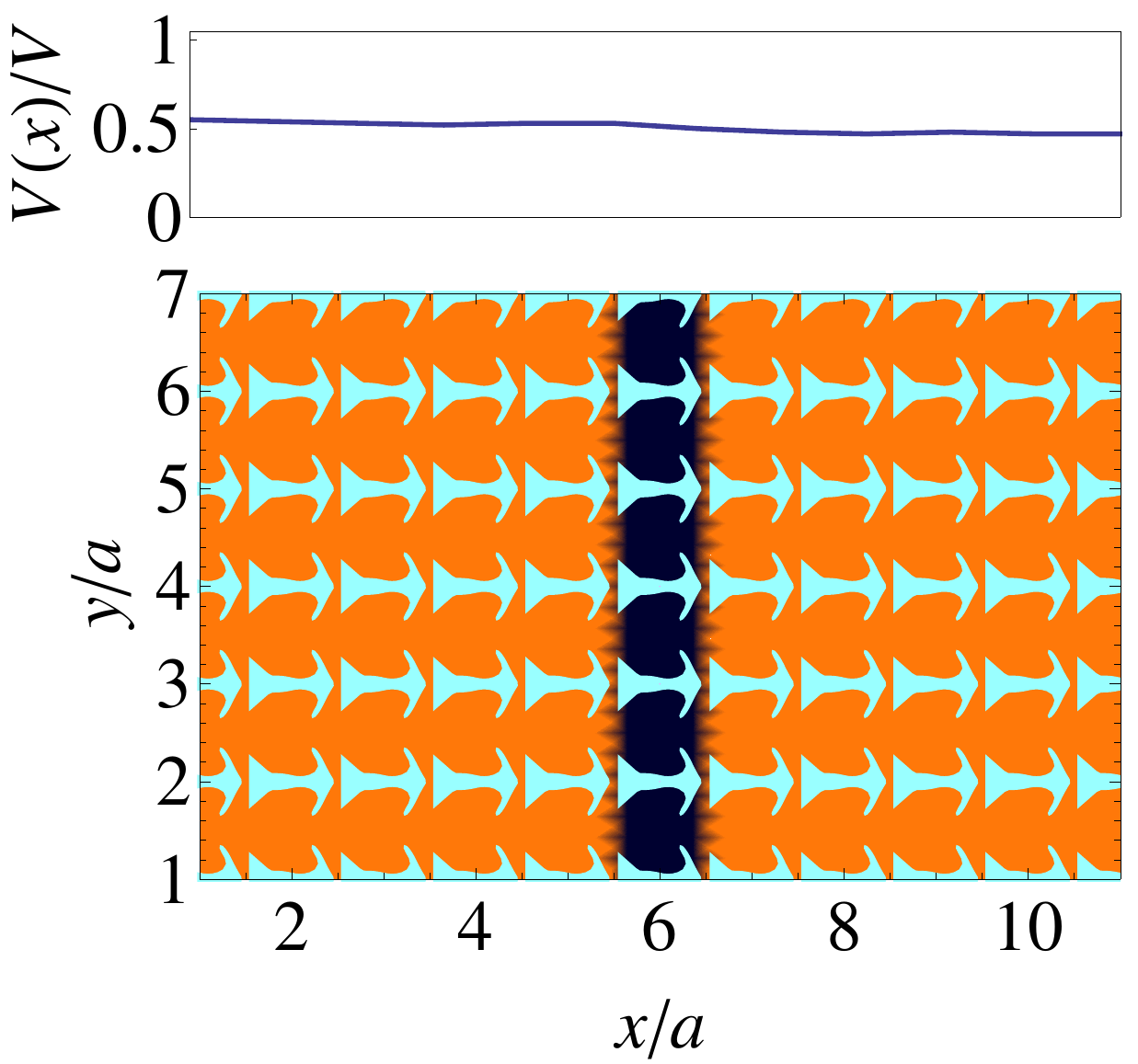}
 \begin{flushleft} (b) Current map for a long barrier $L_{\text{norm}}=4a$
 \end{flushleft}
 \includegraphics[width=0.55\columnwidth]{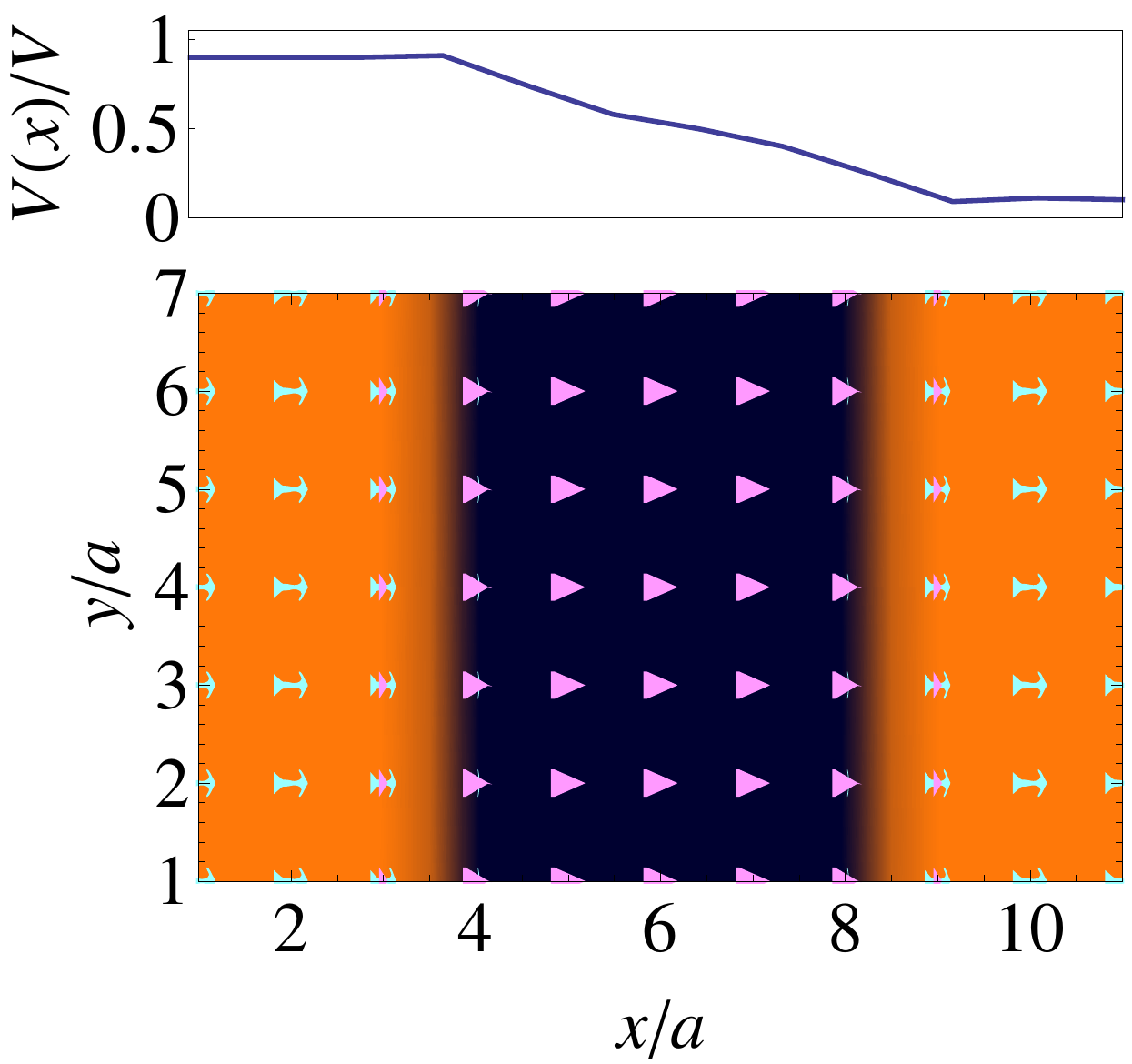}
 \vspace{5 mm}
 \centerline{\quad\quad\resizebox{0.48\linewidth}{!}{\includegraphics{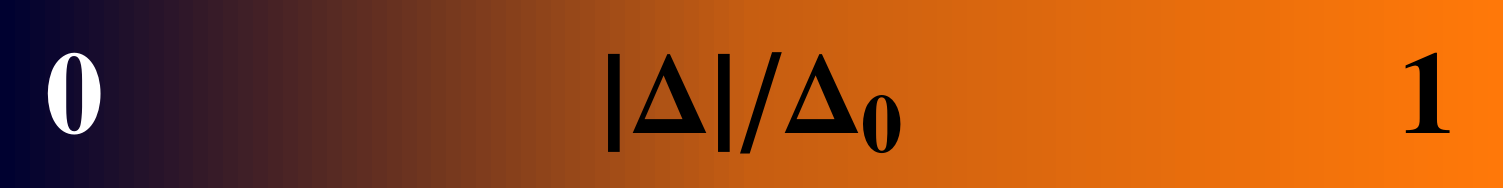}}}
 \caption{(Color online) The upper panel
 shows the potential difference $V(x)$ across the sample with total potential drop $V$.  The lower panel
 shows current maps for short (a) and long barriers (b) respectively. 
 Supercurrent is shown by cyan darts and normal
 current by violet pointers,
 arrow length corresponds to current magnitude and orientation to the
 direction of current flow. Color density corresponds to the order
 parameter $|\Delta|$, which has peak value $\Delta_{0}$.}
 \label{fig:SCNSC}
\end{figure}

\begin{figure}
 \begin{flushleft} (a) Superconductor-insulator transition\end{flushleft}
 \includegraphics[width=0.8\columnwidth]{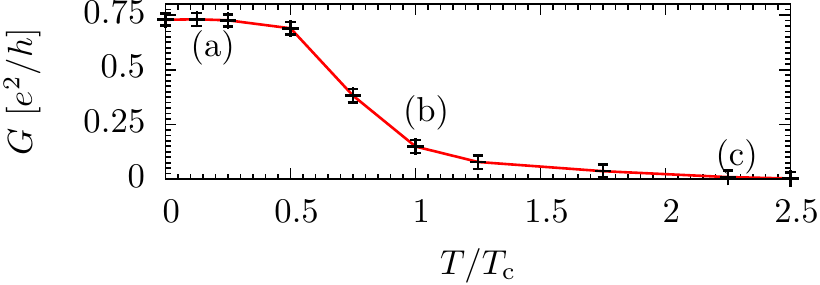}
 \begin{flushleft} (b) SC side of transition, $T\approx0.14T_{\text{c}}$\end{flushleft}
 \includegraphics[width=0.55\columnwidth]{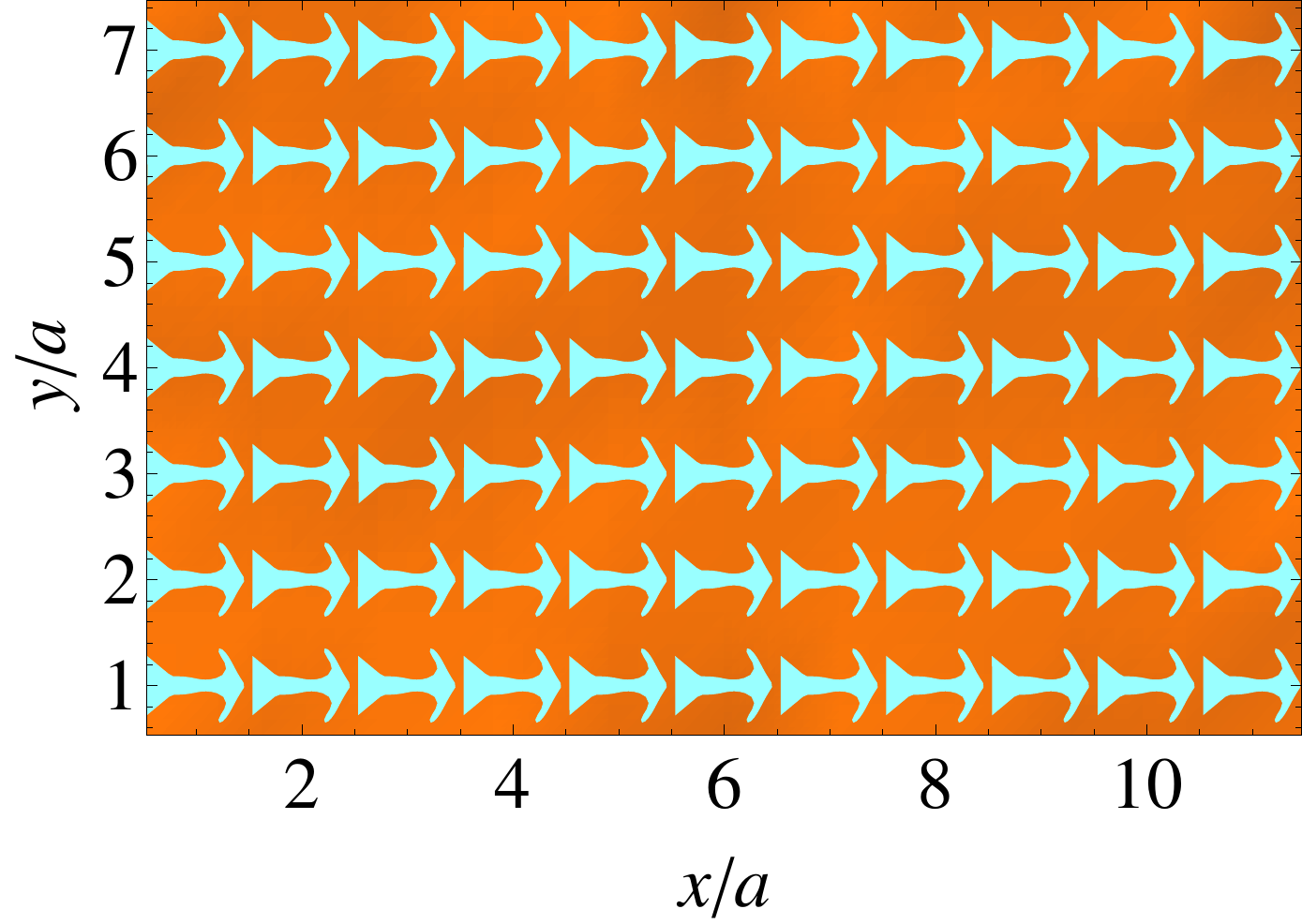}
 \begin{flushleft} (c) At the superconductor-insulator transition,
 $T\approx T_{\text{c}}$ \end{flushleft}
 \includegraphics[width=0.55\columnwidth]{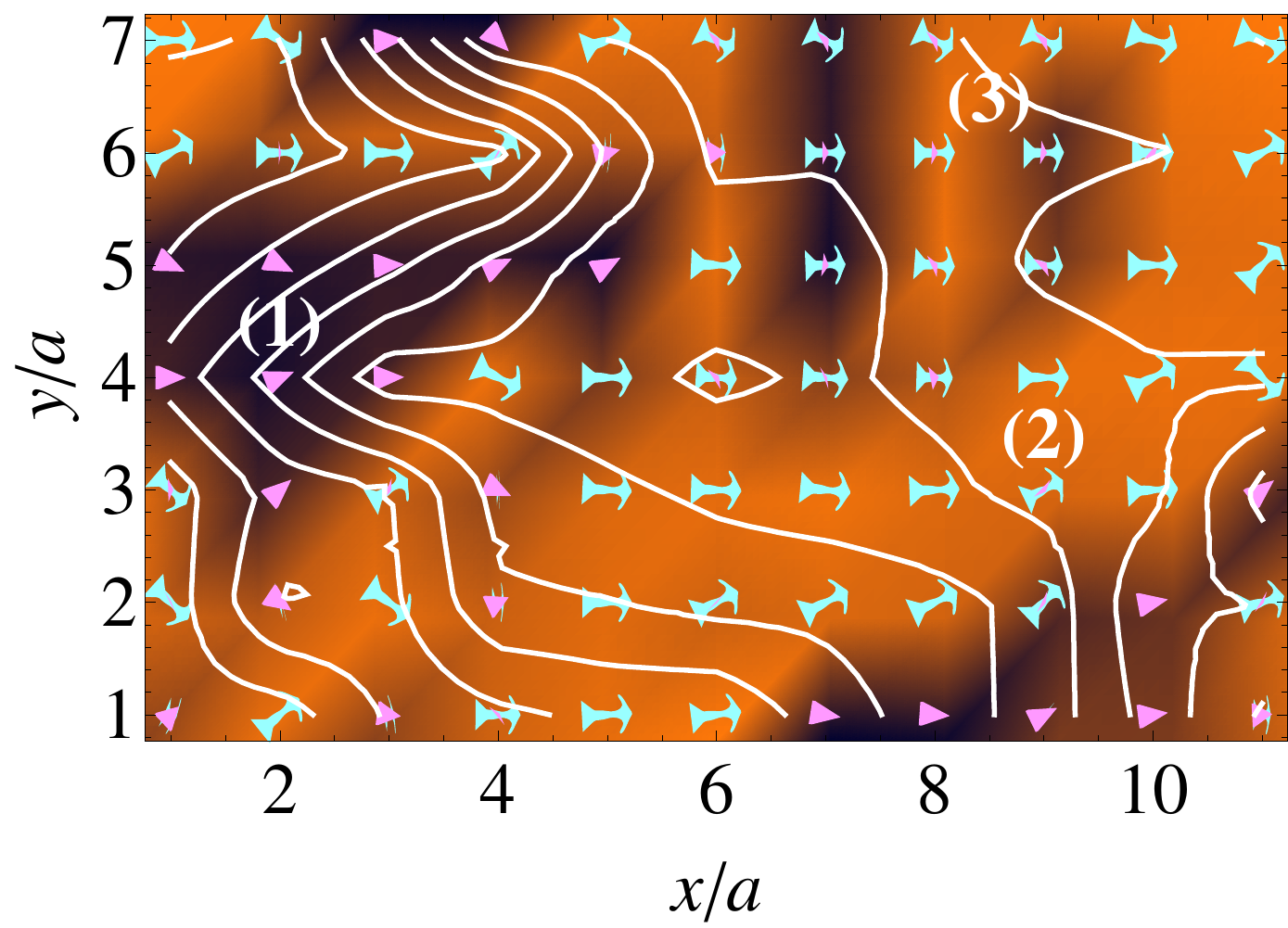}
 \begin{flushleft} (d) Insulating side of transition, $T\approx2.3T_{\text{c}}$\end{flushleft}
 \includegraphics[width=0.55\columnwidth]{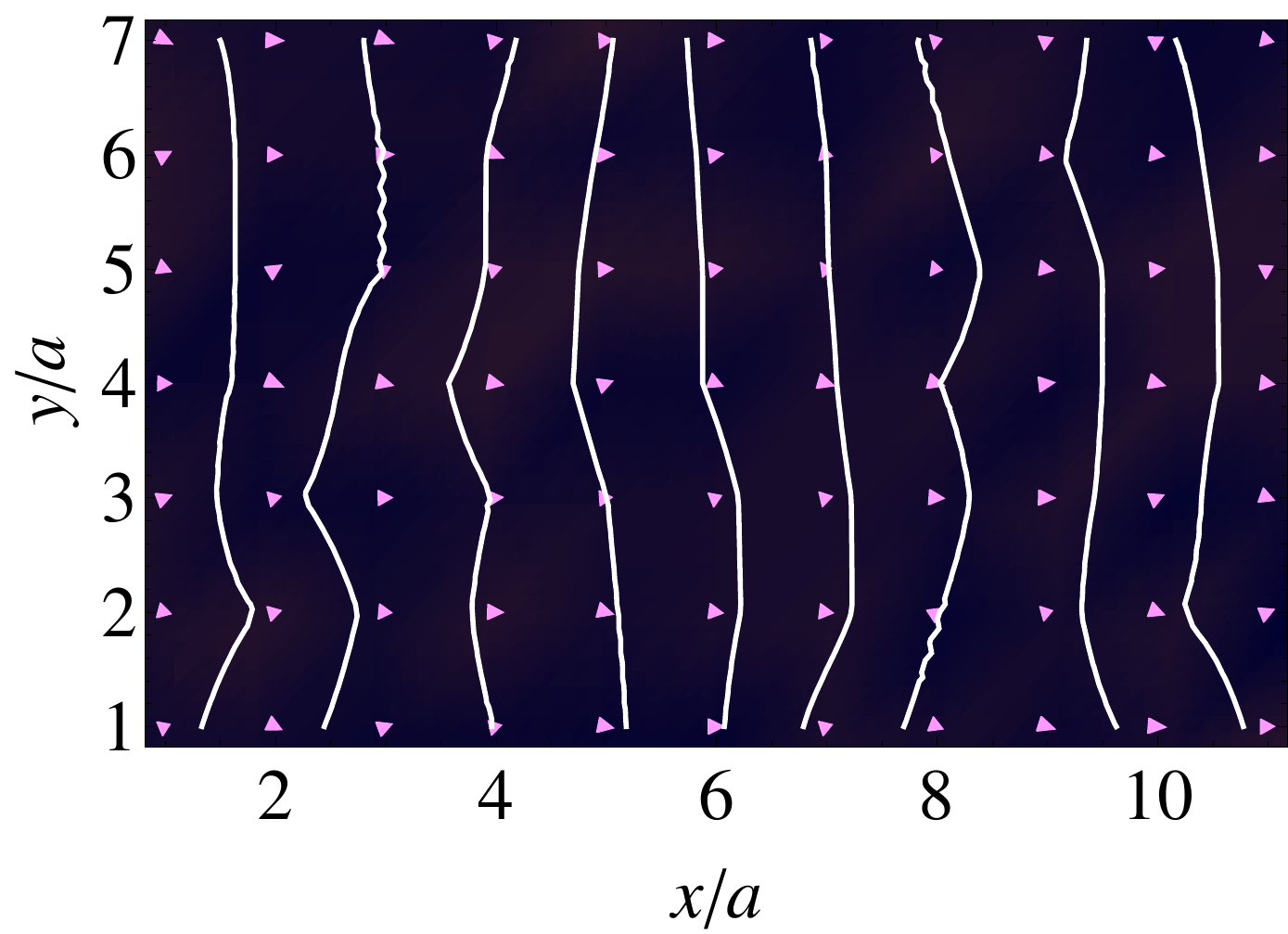}
 \vspace{5 mm}
 \centerline{\quad\quad\resizebox{0.48\linewidth}{!}{\includegraphics{ColorBar.pdf}}}
 \caption{(Color online) (a) Shows the fall in conductance across the superconductor-insulator
 transition. Current maps on tuning temperature from (b) a
 superconductor at $T\approx0.14T_{\text{c}}$ through to (d) an insulator at
 $T\approx2.3T_{\text{c}}$. At $T\approx T_{\text{c}}$ the
 superconductor-insulator transition takes place. Supercurrent is shown
 by cyan darts and normal current by violet pointers, arrow length
 corresponds to current magnitude and orientation to the direction of
 current flow. Color density corresponds to the order parameter
 $|\Delta|$. Lines of equal chemical potential are shown in white. In
 the current map (b) three points of interest are labeled: (1) the
 normal state, (2) the superconductor state, and (3) Josephson tunneling.}
 \label{fig:SCI}
\end{figure}

One important feature of our formalism is the new capability to map
out the flow of both super and normal currents within a sample and the
changes in chemical potential which drive that flow. Since we can now
study the current flow around impurities in the sample and expose weak
links with large potential drop, we should be able to probe phenomena
in the disordered superconductor with unprecedented detail and trace
their cause back to a microscopic mechanism.  While applications of
this formalism to the outstanding problems in this field will be
described in future publications, in this section we aim to
demonstrate the usefulness of the current and potential maps, first by
further studying the Josephson junction with a superconductor
containing a central normal region, and secondly by studying the
superconductor-insulator transition in disordered systems.  However, we
will first verify our current mapping formalism by examining the
site-by-site current conservation in a $39\%$ filled system with no
disorder. As the only sources and sinks of current are the two
metallic leads, a consistent calculation should obey charge
conservation for all of the inner sites of the sample. In
\figref{fig:CurrentConservation}(a) we show the average fractional error in
conservation of current $\sum_{i=1}^{N}|\Delta J_{i}|/J_{i}N$ on each
site as we vary the number of states $K$ included in
the calculation out of a possible $N$ states, as prescribed
in the penultimate paragraph of App.~\ref{MCsection}. We see that
if only $5\%$ of states are included there is a $20\%$ average leakage
of the current. However, if we include $50\%$ of the states in the
calculation of the current there is a leakage of only
$\sim2\%$. Throughout the remainder of this section we include $40\%$
of the states in the calculation to yield an average error of
approximately $3\%$.

Having verified the conservation of current, we demonstrate what can
be learned from the current maps by first studying a modified
Josephson setup consisting of two clean $39\%$ filled SC regions with a
central normal region that has $U=0$. We can then monitor the current
flow through the system to see it change from SC to normal in
character as the intermediate normal region is widened in
\figref{fig:CurrentConservation}(b). For a narrow $U=0$ central
region the two SC regions are phase locked and predominantly a
Josephson current flows (lower panel in \figref{fig:SCNSC}(a)). Due to
the strong proximity effect, the system is entirely SC with no
reduction in conductance. The electrical potential is dropped on the
two contact barriers, and remain constant through the superconductor
(upper panel in \figref{fig:SCNSC}(a)).  (For the present case of two
equal contact barriers the potential in the SC is equal to the average
of the chemical potential of the two leads). On the other hand, when
the central $U=0$ region is wide, $L_{\text{norm}}\gtrsim4a$, the two SC regions are
too weakly coupled for a Josephson current to flow, and instead a
normal current flows between the two SC regions (lower panel in
\figref{fig:SCNSC}(b)). This, in turn, introduces a new resistor
 into the sample and the conductance drops
accordingly. Now the potential drop is mostly across the Josephson
junction (upper panel in \figref{fig:SCNSC}(b)) -- the left-hand
superconductor adopts, approximately, the potential of the left-hand
lead and the right-hand superconductor that of the right-hand
lead. This situation is analogous to current flowing between SC grains
in a disordered sample, and can reveal whether they are coherently
coupled, when a supercurrent flows between the grains, or decoupled,
when a normal current flows. Such analysis could be a vital component
in the study of the origin of resistance in disordered SC system,
and will be used in a subsequent publication, to study the anomalous
magnetoresistance observed in experiment~\cite{Sambandamurthy04}.

We give a glimpse of such an analysis in the case of the
superconductor-insulator transition in a disordered superconductor
with increasing temperature. We take a $39\%$ filled model with weak
disorder, set to $W=0.2t$, which displays a superconductor-insulator
transition at a temperature $T_{\text{c}}\approx0.14t$. In \figref{fig:SCI}(a)
we show the variation of conductance
across the superconductor-insulator transition, and below it study the current
distribution maps. In
\figref{fig:SCI}(b) at $T\approx0.14T_{\text{c}}$ there are weak-disorder driven
fluctuations in the SC order parameter, but an almost uniform
supercurrent. The potential drops mainly in the contacts, and in the
sample is equal to the average of the two leads with small random
fluctuations. In \figref{fig:SCI}(d) at $T\approx2.3T_{\text{c}}$ the
SC order parameter practically vanishes, there is no supercurrent, and,
due to the increasing resistance, only a small normal current flows
through the sample. The potential, as expected for normal systems,
decays linearly across the sample. At intermediate temperatures
$T\approx T_{\text{c}}$ the current map \figref{fig:SCI}(c) highlights
the interplay of the normal and SC current. There is a rough
correlation between regions of finite SC order parameter and supercurrent
flow, on one hand, and zero SC order  parameter and normal current, on the
other.  We point out three typical regions of the sample. Firstly, at
(1) the order parameter is small and only normal current flows,
whereas at (2) the order parameter is large and supercurrent
flows. However, at (3) two SC regions are separated by a small
normal region but are Josephson coupled and so a supercurrent
flows through the zero SC order region. By examining the potential lines we see that the normal
regions, for example (1), are acting as weak links whereas the
potential drop over the superconducting regions is small. Thus the
overall resistance of the sample is dominated by such weak links. The
current and potential maps allow us to see the
superconductor-insulator transition developing, and we plan to
investigate in details the relation of such a percolative picture to
the Kosterlitz-Thouless transition, as was recently suggested
\cite{Erez2010}.

\section{Discussion}

In this paper we have developed a new exact formula to calculate the
current through a superconductor connected to two non-interacting
metallic leads with an imposed potential difference. The formula was
implemented with a negative-$U$ Hubbard model which included both
phase and amplitude fluctuations in the SC order parameter.  A new
Chebyshev expansion method allowed us to solve the model and calculate
the current in $\mathcal{O}(N^{1.9}M^{2/3})$ time, granting access to
systems of unprecedented size. The formalism also enables the
generation of current and potential maps which show exactly where the
super current and separately the normal current flows through the
system.

The formalism was exhaustively tested against a series of
well-established results, demonstrating the accuracy of the procedure,
its ability to capture various physical processes relevant to
superconductivity in disordered systems, and correctly model the
presence of a magnetic field and finite temperature.  These tests
indicate that the formalism and accompanying numerical solver can
robustly calculate the current through a superconductor across a wide
range of systems. In the future we plan to report on the application
of the formalism to several outstanding questions, such as the
magneto-resistance anomaly on crossing the superconductor-insulator
transition~\cite{Sambandamurthy04}, the Little Parks effect in nano-scale cylinders~\cite{Liu01},
 and dissipation-driven phase transitions in SC
wires~\cite{Lobos09}.

{\it Acknowledgments:} GJC acknowledges the financial support of the
Royal Commission for the Exhibition of 1851, the Kreitman Foundation,
and National Science Foundation Grant No. NSF PHY05-51164.  This
work was also supported by the ISF.

\appendix

\section{Derivation of the Current Formula}\label{deriv}

The formula for the current in the Bogoliubov basis set is
\begin{widetext}
\begin{align}
  J\!=\!\frac{\cmplxi e}{2h}\!\sum_{\sigma}\!\int\!\!\diffd\epsilon\Bigl(&\!\!\tr\!\left\{\!\left[f_{\text{L}}(\epsilon)\mat{\Gamma}^{\text{L}}\!-\!f_{\text{R}}(\epsilon)\mat{\Gamma}^{\text{R}}\right]\!\left[\vec{u}_{i}\!\left(\mat{G}_{\sigma}^{>}\!-\!\mat{G}_{\sigma}^{<}\right)\!\vec{u}_{j}^{*}\!+\!\vec{v}_{i}\!\left(\mat{G}_{-\sigma}^{>}\!-\!\mat{G}_{-\sigma}^{<}\right)\!\vec{v}_{j}^{*}\! -\!\sigma\vec{v}_{i}^{*}\!\left(\mat{H}_{\sigma}^{>}\!-\!\mat{H}_{\sigma}^{<}\right)\!\vec{u}_{j}^{*}\!+\!\sigma\vec{u}_{i}\!\left(\bar{\mat{H}}_{-\sigma}^{>}\!-\!\bar{\mat{H}}_{-\sigma}^{<}\right)\!\vec{v_{j}}\right]\!\right\}\neweqnline
  +&\tr\left\{\left[\mat{\Gamma}^{\text{L}}-\mat{\Gamma}^{\text{R}}\right]\left[\vec{u}_{j}^{*}\mat{G}^{<}_{\sigma}\vec{u}_{i}^{*}-\vec{v}_{j}\mat{G}^{>}_{-\sigma}\vec{v}_{i}^{*}
+\sigma\vec{u}^{*}_{j}\mat{H}_{\sigma}^{>}\vec{v}^{*}_{i}-\sigma\vec{v}_{j}\bar{\mat{H}}_{-\sigma}^{<}\vec{u}_{i}\right]\right\}\Bigr)\punc{.}
\label{eqn:CurrentInLesserGreenFns}
\end{align}

We need to determine the Green functions across the sample, which must
be calculated in the presence of the leads. However, as the electrons
in the metallic leads are non-interacting we can start from the bare
electronic Green functions for the superconductor not coupled to the
leads
$\tilde{G}^{\text{r}}_{\text{e}\sigma}(m,n)=\delta_{m,n}/(\epsilon-\xi_{m}+\cmplxi\delta)$
and
$\tilde{G}^{\text{r}}_{\text{h}\sigma}(m,n)=\delta_{m,n}/(\epsilon+\xi_{m}+\cmplxi\delta)$,
which have energy eigenstates $\xi_{m}$ and $\delta\to0^{+}$. We then
write down Dyson's equation to self-consistently include the leads
\begin{align}
 \left(
 \begin{array}{c}
  \mat{G}^{\text{r}}_{\sigma}\\\mat{H}^{\text{r}}_{\sigma}
 \end{array}
 \right)
 =
 \left(
 \begin{array}{c}
  \mat{\tilde{G}}^{\text{r}}_{\text{e}\sigma}\\0
 \end{array}
 \right)
 +
 V^{2}\left(
 \begin{array}{cc}
  \mat{\tilde{G}}^{\text{r}}_{\text{e}\sigma}(\vec{u}_{\vec{p}}^{*}g^{\text{r}}_{\text{e}\vec{p}\chi}\vec{u}_{\vec{p}}+\vec{v}_{\vec{p}}^{*}g^{\text{r}}_{\text{h}\vec{p}\chi}\vec{v}_{\vec{p}})
  &\sigma\mat{\tilde{G}}^{\text{r}}_{\text{e}\sigma}(\vec{v}^{*}_{\vec{p}}g^{\text{r}}_{\text{h}\vec{p}\chi}\vec{u}_{\vec{p}}^{*}-\vec{u}_{\vec{p}}^{*}g^{\text{r}}_{\text{e}\vec{p}\chi}\vec{v}^{*}_{\vec{p}})\\
  \sigma\mat{\tilde{G}}^{\text{r}}_{\text{h}\sigma}(\vec{u}_{\vec{p}}g^{\text{r}}_{\text{h}\vec{p}\chi}\vec{v}_{\vec{p}}-\vec{v}_{\vec{p}}g^{\text{r}}_{\text{e}\vec{p}\chi}\vec{u}_{\vec{p}})
  &\mat{\tilde{G}}^{\text{r}}_{\text{h}\sigma}(\vec{u}_{\vec{p}}g^{\text{r}}_{\text{h}\vec{p}\chi}\vec{u}_{\vec{p}}^{*}+\vec{v}_{\vec{p}}g^{\text{r}}_{\text{e}\vec{p}\chi}\vec{v}^{*}_{\vec{p}})
 \end{array}
 \right)
 \left(
 \begin{array}{c}
  \mat{G}^{\text{r}}_{\sigma}\\\mat{H}^{\text{r}}_{\sigma}
 \end{array}
 \right)\punc{.}
 \label{eqn:MatrixEquationForRetardedAndAdvancedGreensFunctions}
\end{align}
\end{widetext}
Here
$g_{\text{e}\vec{p}\chi}^{\text{r}}=1/(\epsilon-\varepsilon_{\vec{p}}+\mu_{\chi}+\cmplxi\delta)$
is the retarded Green function of the non-interacting electrons in the
leads, with dispersion $\varepsilon_{\vec{p}}$, and
$\{\mat{u}_{\vec{p}},\mat{v}_{\vec{p}}\}$ are the matrices of the
eigenstates multiplied by the lead plane wave states $\vec{p}$ at the
tunneling barriers. To extract the retarded Green function and its
anomalous counterpart from this matrix equation one has to perform a
matrix inversion. The Dyson equation is for the retarded and advanced
Green functions, whereas the current formula
\eqnref{eqn:CurrentInLesserGreenFns} is in terms of the lesser and
greater Green functions. To transform these into the retarded and
advanced Green functions we apply the identity
$\mat{G}_{\sigma}^{<}=\tilde{\mat{G}}_{\sigma}^{<}+\tilde{\mat{G}}_{\sigma}^{\text{r}}\mat{\Sigma}_{\sigma}^{\text{r}}\mat{G}_{\sigma}^{<}+\tilde{\mat{G}}_{\sigma}^{\text{r}}\mat{\Sigma}_{\sigma}^{<}\mat{G}_{\sigma}^{\text{a}}+\tilde{\mat{G}}_{\sigma}^{<}\mat{\Sigma}_{\sigma}^{\text{r}}\mat{G}_{\sigma}^{\text{a}}$
recursively to find
$\mat{G}_{\sigma}^{<}=(1+\mat{G}_{\sigma}^{\text{r}}\mat{\Sigma}_{\sigma}^{\text{r}})\tilde{\mat{G}}_{\sigma}^{<}(1+\mat{\Sigma}_{\sigma}^{\text{a}}\mat{G}_{\sigma}^{\text{a}})+\mat{G}_{\sigma}^{\text{r}}\mat{\Sigma}_{\sigma}^{<}\mat{G}_{\sigma}^{\text{a}}$,
where $\mat{\Sigma}_{\sigma}$ is the self energy.  This recursion
fixes the chemical potential of the superconductor by including
tunneling to and from the leads. This will ensure that the net number
of electrons is conserved, analogous to some extensions to the BTK
formalism~\cite{Lambert91}. However, as the final chemical potential
must be independent of the chemical potential of the uncoupled
superconductor, the term containing $\tilde{\mat{G}}^{<}$ must be
identically zero leaving
$\mat{G}^{<}_{\sigma}=\mat{G}^{\text{r}}_{\sigma}\mat{\Sigma}^{<}_{\sigma}\mat{G}^{\text{a}}_{\sigma}$,
and its greater Green function counterpart
$\mat{G}^{>}_{\sigma}=\mat{G}^{\text{r}}_{\sigma}\mat{\Sigma}^{>}_{\sigma}\mat{G}^{\text{a}}_{\sigma}$.
We now extend this identity to include the anomalous Green function
and recover
\begin{widetext}
\begin{align}
 \left(
 \begin{array}{c}
  \mat{G}^{<}_{\sigma}\\\mat{H}^{<}_{\sigma}
 \end{array}
 \right)
 =
 V^{2}\left(
 \begin{array}{cc}
  \mat{G}^{\text{r}}_{\sigma}(\vec{u}_{\vec{p}}^{*}g^{<}_{\text{e}\vec{p}\chi}\vec{u}_{\vec{p}}+\vec{v}^{*}_{\vec{p}}g^{<}_{\text{h}\vec{p}\chi}\vec{v}_{\vec{p}})
  &\sigma\mat{G}^{\text{r}}_{\sigma}(\vec{v}^{*}_{\vec{p}}g^{<}_{\text{h}\vec{p}\chi}\vec{u}^{*}_{\vec{p}}-\vec{u}^{*}_{\vec{p}}g^{<}_{\text{e}\vec{p}\chi}\vec{v}^{*}_{\vec{p}})\\
  \sigma\mat{H}^{\text{r}}_{\sigma}(\vec{u}_{\vec{p}}g^{<}_{\text{h}\vec{p}\chi}\vec{v}_{\vec{p}}-\vec{v}_{\vec{p}}g^{<}_{\text{e}\vec{p}\chi}\vec{u}_{\vec{p}})
 &\mat{H}^{\text{r}}_{\sigma}(\vec{u}_{\vec{p}}g^{<}_{\text{h}\vec{p}\chi}\vec{u}^{*}_{\vec{p}}+\vec{v}_{\vec{p}}g^{<}_{\text{e}\vec{p}\chi}\vec{v}^{*}_{\vec{p}})
 \end{array}
 \right)
 \left(
 \begin{array}{c}
  \mat{G}^{\text{a}}_{\sigma}\\\mat{H}^{\text{a}}_{\sigma}
 \end{array}
 \right)\punc{.}
  \label{eqn:EqnForGLesser}
\end{align}
\end{widetext}
We can now take this, the analogous expression for the greater Green
function, and their anomalous counterparts, and substitute them into
\eqnref{eqn:CurrentInLesserGreenFns}, which will yield \eqnref{eqn:GeneralEquationForCurrent}.

\section{Evaluation of the Monte Carlo Integrals}
\label{MCsection}

\begin{figure}
 \includegraphics[width=0.9\columnwidth]{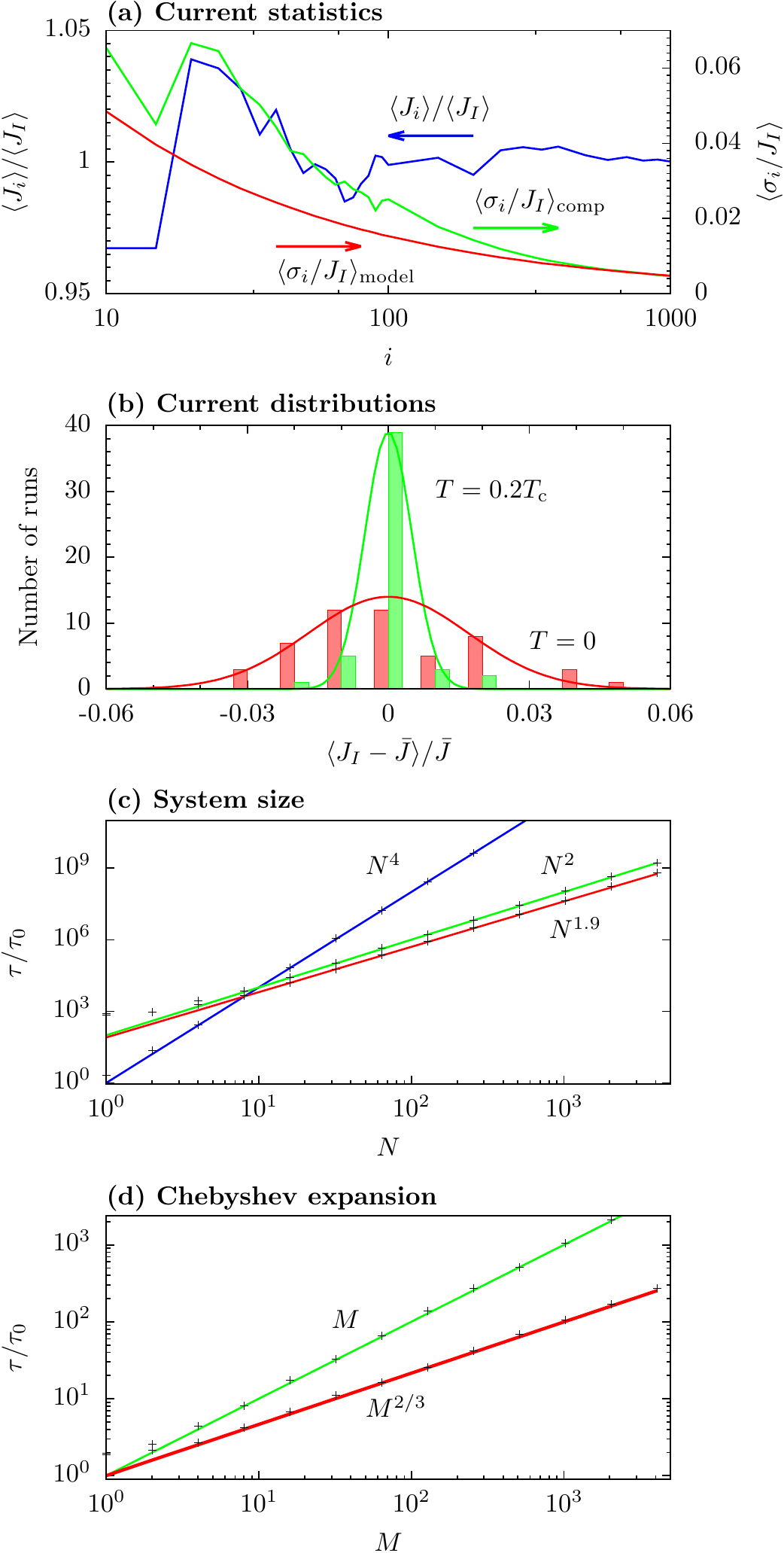}
 \caption{(Color online) (a) The estimate of the current with number
 of Monte Carlo iterations, $i$, out of a total number $I=1000$. The
 primary y-axis shows the best estimate of the current (blue). The
 secondary y-axis shows the estimated standard deviation in this
 estimate (green) and idealized improvement in the accuracy (red). (b)
 The distribution of 50 separate current estimates at $T=0$ (red)
 and $T=0.2T_{\text{c}}$ (green) with best-fit Gaussian distributions. (c and d)
 The time $\tau$ to perform a run on a $32\times32$ system
 renormalized by the time $\tau_{0}$ for a $M=512$, $N=1$ system. In
 (c) the change with varying the system size $N$, where the blue line
 is for the standard $\mathcal{O}(N^{4})$ method of finding all of the
 energy eigenvalues, the green is the $\mathcal{O}(N^{2})$ standard
 Chebyshev expansion method~\cite{Weisse09}, and the blue is the
 $\mathcal{O}(N^{1.9})$ extended Chebyshev approach. In (d) the two
 Chebyshev expansion method approaches are compared by varying the
 expansion order $M$. As in (c), the green line is the standard
 $\mathcal{O}(M)$ approach~\cite{Weisse09}, and the red line is the
 new $\mathcal{O}(M^{2/3})$ algorithm.}  \label{fig:CalcTimePlot}
\end{figure}

In order to evaluate the correlation functions
(e.g. Eq.~\ref{eq:rhoO}), we need to sum over all possible spatial
configurations of the auxiliary fields $\vecgrk{\rho}$ and
$\{\vecgrk{\Delta},\bar{\vecgrk{\Delta}}\}$, with each configuration
carrying the weight $P(\vecgrk{\rho},\vecgrk{\Delta})=\exp(-\beta
E[\vecgrk{\rho},\vecgrk{\Delta}])/\mathcal{Z}$. This distribution is sampled
using the Metropolis algorithm~\cite{Metropolis53}, which at each step
proposes a new configuration of either the field $\vecgrk{\rho}$ or
$\vecgrk{\Delta}$ and calculates the resulting change in the total
energy. If this change in the energy is negative the step is accepted,
whereas if positive it is accepted with probability
$\exp\{-\beta(E[\vecgrk{\rho}_{\text{new}}]-E[\vecgrk{\rho}_{\text{old}}])\}$
and
$\exp\{-\beta(E[\vecgrk{\Delta}_{\text{new}}]-E[\vecgrk{\Delta}_{\text{old}}])\}$
respectively. Since the walk over $\vecgrk{\rho}$ is one-dimensional
we choose the step size
$|\vecgrk{\rho}_{\text{new}}-\vecgrk{\rho}_{\text{old}}|$ to aim for
$50\%$ of the steps to be accepted, whereas the walk over
$\{\vecgrk{\Delta},\bar{\vecgrk{\Delta}}\}$ covers a two-dimensional
space so we choose a step size
$|\vecgrk{\Delta}_{\text{new}}-\vecgrk{\Delta}_{\text{old}}|$ so that
$35.2\%$ of the steps will be accepted~\cite{Gelman96}.

Central to the Monte Carlo method used to sample the partition
function is the requirement to calculate the energy difference between
two different configurations of the auxiliary fields,
$\{\vecgrk{\rho}_{\text{old}},\vecgrk{\Delta}_{\text{old}}\}$ and
$\{\vecgrk{\rho}_{\text{new}},\vecgrk{\Delta}_{\text{new}}\}$. For a
lattice with $N$ sites, to calculate the energy of each proposed
configuration requires an effort of $\mathcal{O}(N^{3})$, so an entire
sweep over the $N$ sites that make up the fields $\vecgrk{\rho}$ and
$\{\vecgrk{\Delta},\bar{\vecgrk{\Delta}}\}$ requires a computational
effort of $\mathcal{O}(N^{4})$. However, a recent method developed by
Wei{\ss}e~\cite{Weisse09} calculates just the difference between the
energy of the configurations in a computationally efficient
manner. For an update to the $i\text{th}$ site a Chebyshev expansion
with the $0\leq m\leq M$ coefficients containing $\langle
i|\mat{T}_{m}(\hat{H}/s)|i\rangle$ must be calculated, where
$\mat{T}_{m}$ is defined by the recursion relation
$\mat{T}_{m}(\mat{x})=2\mat{x}\mat{T}_{m-1}(\mat{x})-\mat{T}_{m-2}(\mat{x})$,
$\mat{T}_{0}(\mat{x})=\mat{I}$, and $\mat{T}_{1}(\mat{x})=\mat{x}$. A
typical expansion contained $M=1024$ terms. Previous
authors~\cite{Weisse09} have calculated this site-by-site through a
succession of sparse matrix-vector multiplications, each of cost
$\mathcal{O}(NM)$, so for an entire sweep over the order parameter the
computational effort is $\mathcal{O}(N^{2}M)$. However, here we
optimize the programme so that the entire sweep can be performed in
$\mathcal{O}(N^{1.9}M^{2/3})$ time. Rather than follow a site-by-site
approach calculated with sparse matrix-vector multiplications we
instead calculate the matrix elements for the entire sweep
simultaneously, which necessitates performing matrix-matrix
multiplications. Provided the changes in the order parameters are
small the local changes are independent of those of surrounding sites
and we can then perform the entire sweep from this data set. Spherical
averaging further reduces the influence of changes in the surrounding
order parameters. Central to the recursion relation for $\mat{T}_{m}$
is the costly calculation of $\mat{x}^{n}$, for $1<n\leq M$. To
evaluate this we divide the calculation of the $M$ matrix products
into three stages:
\begin{enumerate}
 \item The lowest order matrix products, up to $\mat{x}^{k}$, are
 sparse. Therefore, for the elements $1<n\leq k$ the matrix
 multiplications involve only sparse matrices, each of peak cost $kN$,
 and the total cost of calculating them is $\mathcal{O}(k^{2}N)$.
 \item The second stage is to successively calculate every
 $k\text{th}$ matrix product. Each of these involves multiplying the
 dense matrix $\mat{x}^{pk}$ by the matrix $\mat{x}^{k}$, for integer
 $1\leq p\leq M/k$, which costs $\mathcal{O}(N^{2.38})$
 time~\cite{Coppersmith90}. With $M/k$ of these products to calculate
 the total cost is $\mathcal{O}(N^{2.38}M/k)$.  \item The third stage
 is to construct the entire family of $\mat{x}^{n}$ by interpolating
 between the matrices $\mat{x}^{pk}$ found in the second stage. This
 is done by multiplying the dense matrices found in the second stage
 by the sparse matrices found in the first stage. Furthermore, as we
 need only the diagonal elements of the final matrix each separately
 costs $\mathcal{O}(kN)$ and so the total cost is $\mathcal{O}(kNM)$.
\end{enumerate}
Having now laid out the prescription of how to calculate the matrix
elements, we now examine the total cost,
$\mathcal{O}(k^{2}N+N^{2.38}M/k+kNM)$. The choice
$k\sim\sqrt[3]{N^{1.38}M}$ will minimize the total cost to
$\mathcal{O}(N^{1.9}M^{2/3}+N^{1.46}M^{4/3})$, and as typically $N\gg
M$ the cost is $\sim\mathcal{O}(N^{1.9}M^{2/3})$. This is a
significant improvement over the cost $\mathcal{O}(N^{2}M)$ of the
Chebyshev expansion approach~\cite{Weisse09}, which for the parameters
employed in our simulations corresponds to a speedup by a factor of
$\sim30$. Now that the matrix elements behind the Chebyshev expansion
have been found they are applied for the entire sweep.

To verify the Monte Carlo procedure in \figref{fig:CalcTimePlot}(a) we
first check the convergence of the estimate for the current and that
its standard error falls as the root of the number of Monte Carlo
iterations. In \figref{fig:CalcTimePlot}(b) we compare the results of
equilibrated Monte Carlo runs at zero temperature from a variety of
initial configurations of the order parameter fields $\rho$ and
$\Delta$. Evolution under the Metropolis algorithm drives these
starting fields into different relaxed configurations, which because
the simulations are restricted here to $T=0$ are unable to be excited
out to explore different configurations. These final configurations
yield a variety of different current values, with standard deviation
of $\sim\pm2.4\%$ of the final total current. At finite temperature
thermal excitations
can drive the system to explore configurations around the ground state
with a narrower standard deviation of $\sim\pm0.6\%$. Having verified the
current statistics, in \figref{fig:CalcTimePlot}(c and d) we show the
results of some timing runs that highlight the improvement of the
algorithm to $\mathcal{O}(N^{1.9}M^{2/3})$ time over the standard
approach of calculating all the energy eigenvalues in
$\mathcal{O}(N^{4})$ time and the standard Chebyshev approach that
runs in $\mathcal{O}(N^{2}M)$ time. In particular, by varying the
system size we observe that the method of calculating all the
eigenvalues is more efficient for systems smaller than $N\sim10$, but
the new Chebyshev approach is superior for large systems. We took
advantage of this development to study systems of unprecedented size.

The Chebyshev expansion method just described represents a zero order
approximation. However, we can extend this method further and
calculate the lowest order change in the Chebyshev expansion following
a shift in the configuration of the fields $\vecgrk{\rho}$ and
$\vecgrk{\Delta}$ by $\vecgrk{\delta}$. The resultant shift in the
Chebyshev expansion of $\mat{T}_i$ is found using the recursion
relationships $\mat{t}_{i}=\frac{2}{s}\vecgrk{\delta}\mat{T}_{i-1}
+\frac{2}{s}\mat{H}\mat{t}_{i-1}-\mat{t}_{i-2}$ with $\mat{t}_{0}=0$
and $\mat{t}_1=\vecgrk{\delta}/s$. This allows the Chebyshev expansion
coefficients to be extrapolated over several configuration space
sweeps, and the calculation time falls proportionally. Spherical
averaging also reduces the influence of changes in the surrounding
order parameters. In practice it was found that up to ten
extrapolation steps could be performed, resulting in a code speed-up
of a factor of ten.

Though the Chebyshev approach can be used to direct the sampling of
the system, to calculate expectation values, such as the current, it
is necessary to diagonalize the system and determine the field
configurations of its states. Formally this requires
$\mathcal{O}(N^{3})$ time. However, since the current is dominated by
the quasiparticle states near to the Fermi surface we instead adopt
the Implicitly Restarted Arnoldi Method~\cite{Lehoucq96} to calculate
only those particular states. We are also helped by the sparsity of
the matrix, which allows us to calculate $K$ eigenstates in
$\mathcal{O}(KN)$ time. It is usually necessary to calculate a certain
fraction of the energy states, so $K\propto N$, and the total cost is
$\mathcal{O}(N^{2})$. The eigenfunctions and energies can then be used
to calculate the current for a specific realization of $\vecgrk{\rho}$
and $\vecgrk{\Delta}$ using the formalism described in
\secref{sec:AnalyticalFormalism}. It is then necessary to average over
successive realizations of $\vecgrk{\rho}$ and
$\vecgrk{\Delta}$. However, the contribution from successive Monte
Carlo calculations might be serially correlated which would result in
an underestimated value for the uncertainty in the predicted value of
the current. To correct for this we calculated the correlation time
through the truncated autocorrelation
function~\cite{Grotendorst02}. We find a typical correlation time of
approximately six Monte Carlo steps, which without autocorrelation
corrections would correspond to an underestimate in the uncertainty of
a factor of $\sim2.5$.


\end{document}